\documentclass[fleqn,usenatbib]{mnras}

\usepackage[T1]{fontenc}
\usepackage{ae,aecompl}

\usepackage{geometry}
\usepackage{graphicx}	
\usepackage{amsmath}	

\usepackage{siunitx} 
\usepackage{booktabs}

\usepackage{algorithm}
\usepackage[noend]{algpseudocode}

\usepackage{mathrsfs}

\DeclareMathOperator\erf{erf}
\DeclareMathOperator\Si{Si}
\DeclareMathOperator\Ci{Ci}

\newcommand{\dd}[1]{\,\text{d}#1}

\usepackage{xcolor}

\definecolor{arsenic}{rgb}{0.23, 0.27, 0.29}
\definecolor{bostonuniversityred}{rgb}{0.8, 0.0, 0.0}
\definecolor{bostonred}{rgb}{0.8, 0.0, 0.0}
\definecolor{darkgreen}{rgb}{0.0, 0.2, 0.13}
\definecolor{amber}{rgb}{1.0, 0.75, 0.0}
\definecolor{airforceblue}{rgb}{0.36, 0.54, 0.66}
\definecolor{i6peach}{HTML}{FFE9C0}
\definecolor{camouflagegreen}{rgb}{0.47, 0.53, 0.42}
\definecolor{asparagus}{rgb}{0.53, 0.66, 0.42}
\definecolor{i6blue}{cmyk}{1,0.305,0,0.06}
\definecolor{i6carmine}{HTML}{960018} 

\usepackage{tikz}
\usetikzlibrary{shapes.geometric, shapes.multipart, arrows, decorations.markings}
\usepackage{varwidth}
\usetikzlibrary{arrows}
\tikzstyle{data}=[rectangle split,rectangle split parts=2,draw,text centered]

\tikzstyle{buildblocksplit} = [ rectangle, 
                                rounded corners, 
                                minimum width=1cm,
                                minimum height=1cm,
                                text centered, 
                                draw=black,
                                rectangle split,
                                rectangle split parts=2,
                                rectangle split part fill={bostonuniversityred!50,
                                                           i6peach} ]
\tikzstyle{buildblock} = [ rectangle, 
                           rounded corners, 
                           minimum width=1cm,
                           minimum height=1cm,
                           text centered, 
                           draw=black,
                           fill=bostonuniversityred!50 ]
\tikzstyle{outblock} = [ trapezium, 
                         trapezium left angle=80, 
                         trapezium right angle=100, 
                         minimum width=1cm, 
                         minimum height=1cm, 
                         text centered, 
                         draw=black, 
                         fill=i6peach, 
                         execute at begin node=\begin{varwidth}{2cm}\begin{center}, execute at end node=\end{center}\end{varwidth}]
\tikzstyle{inblock} = [ rectangle, 
                        rounded corners, 
                        minimum width=1cm,
                        minimum height=1cm,
                        text centered, 
                        draw=black, 
                        fill=i6blue!30]
\tikzstyle{inoutblock} = [ trapezium, 
                           trapezium left angle=80, 
                           trapezium right angle=100,
                           rounded corners, 
                           minimum width=1cm,
                           minimum height=1cm,
                           text centered, 
                           draw=black, 
                           fill=i6blue!30, 
                           execute at begin node=\begin{varwidth}{2cm}\begin{center}, 
                           execute at end node=\end{center}\end{varwidth} ]
\tikzstyle{process} = [ circle, 
                        minimum width=1cm, 
                        minimum height=1cm, 
                        text centered, 
                        draw=black, 
                        fill=airforceblue!30, 
                        execute at begin node=\begin{varwidth}{7cm}\begin{center}, 
                        execute at end node=\end{center}\end{varwidth}]
\tikzstyle{decision} = [ rectangle, 
                         minimum width=1cm, 
                         minimum height=1cm, 
                         text centered, 
                         draw=black, 
                         fill=asparagus!70]
\tikzstyle{arrow} = [thick,->,line width=1pt, i6carmine]
\tikzstyle{dashedarrow} = [thick,dashed,->,line width=1pt, i6carmine]
\tikzstyle{vecArrow} = [thick, i6carmine, decoration={markings,mark=at position
   1 with {\arrow[thick, i6carmine]{open triangle 60}}},
   double distance=1.4pt, shorten >= 5.5pt,
   preaction = {decorate},
   postaction = {draw,line width=1.4pt, white,shorten >= 4.5pt}]
\tikzstyle{innerWhite} = [semithick, white,line width=1.4pt, shorten >= 4.5pt]

\usepackage{newtxtext,newtxmath}

\title[Introducing ScamPy]{ScamPy -- A sub-halo clustering \& abundance matching based Python interface for painting galaxies on the dark matter halo/sub-halo hierarchy}

\author[T. Ronconi et al.]{
Tommaso Ronconi,$^{1,2,3}$\thanks{E-mail: \href{mailto:tronconi@sissa.it}{tronconi@sissa.it} (SISSA)}
Andrea Lapi,$^{1,2,3,4}$
Matteo Viel,$^{1,2,3,4}$
and Alberto Sartori$^{1}$
\\
$^{1}$SISSA, Via Bonomea 265, 34136 Trieste, Italy\\
$^{2}$IFPU, Via Beirut 2, 34014 Trieste, Italy\\
$^{3}$INFN-Sezione di Trieste, via Valerio 2, 34127 Trieste,  Italy\\
$^{4}$INAF-OATS, via Tiepolo 11, 34131 Trieste, Italy\\
}

\date{Accepted XXX. Received YYY; in original form ZZZ}

\pubyear{XXXX}

\begin{document}
\label{firstpage}
\pagerange{\pageref{firstpage}--\pageref{lastpage}}
\maketitle

\begin{abstract}
We present a computational framework for ``painting'' galaxies on top of the Dark Matter Halo/Sub-Halo hierarchy obtained from N-body simulations. 
The method we use is based on the sub-halo clustering and abundance matching (SCAM) scheme which requires observations of the 1- and 2-point statistics of the target (observed) population we want to reproduce.
This method is particularly tailored for high redshift studies and thereby relies
on the observed high-redshift galaxy luminosity functions and correlation properties. 
The core functionalities are written in c++ and exploit Object Oriented Programming, with a wide use of polymorphism, to achieve flexibility and high computational efficiency.
In order to have an easily accessible interface, all the libraries are wrapped in python and provided with an extensive documentation.
We validate our results and provide a simple and quantitative application to 
reionization, with an investigation of physical quantities related to the
galaxy population, ionization fraction and bubble size distribution.
The library is publicly available at \href{https://github.com/TommasoRonconi/scampy}{https://github.com/TommasoRonconi/scampy} with full documentation and examples at \href{https://scampy.readthedocs.io}{https://scampy.readthedocs.io}.
\end{abstract}





\section{Introduction}
\label{sec:intro}

Cosmological N-body simulations are a fundamental tool for assessing the non-linear evolution of the large scale structure (LSS).
With the increasing power of computational facilities, cosmological N-body simulations have grown in size and resolution, allowing to study extensively the formation and evolution of dark matter (DM) haloes \citep{springel2005,Boylan-Kolchin2009,Klypin2011,angulo2012,klypin2016}.
Our confidence on the reliability of these simulations stands on the argument that the evolution of the non-collisional matter component only depends on the effect of gravity and on the initial conditions.
While for the first, we can rely on a solid theoretical background, with analytical solutions for both the classical gravitation theory and for a wide range of its modifications, for the latter, we have measurements at high accuracy \citep{Planck2018} of the primordial power spectrum of density fluctuations.

The formation and evolution of the luminous component (i.e. galaxies and intergalactic baryonic matter) are far from being understood at the same level as the dark matter.
Several possible approaches have been attempted so far to asses this modeling issue, which can be divided into two main categories.
On one side, \textit{ab initio} models, such as N-body simulations with a full hydrodynamical treatment and semi-analytical models, that should incorporate all the relevant astrophysical processes, are capable of tracing back the evolution in time of galaxies within their DM host haloes \citep[see][for reviews]{sommerville-dave2015,naab-ostriker2017}.

On the other side, \textit{empirical} (or \textit{phenomenological}) models are designed to reproduce observable properties of a target (observed) population of objects at a given moment of their evolution \citep[see, e.g., ][ for a review]{Wechsler2018}.
This latter class of methods is typically cheaper in terms of computational power and time required for running.
The development of an approach to model the luminous component without prior assumptions on the baryon physics have emerged from the advent of large galaxy surveys \citep{York2000,colles2001,Lilly2007,driver2011,Grogin2011,mccracken2012}.

Building an empirical model of galaxy occupation requires to define the hosted-object/hosting-halo connection for associating to the underlying DM distribution its baryonic counterpart.
This has been achieved by exploiting several approaches that span from the classical mass-based methods, such as the Halo Occupation Distribution (HOD) scheme \citep{peacock2000,seljak2000,white2001,berlind2002,Yang2003,Zehavi2004,Zheng2005,Tinker2005,Brown2008,Leauthaud2012} or the sub-halo abundance matching (SHAM) scheme \citep{mo1996,Wechsler1998,vale2004,Conroy2006,wang2006,wang2007,Moster2010,Behroozi2010,guo2010,Trujillo_Gomez_2011}, to more sophisticated parameterisations that follow the halo evolution in time \citep{conroy2009,yang2012,behroozi2013,moster2013,moster2018,zhu2020} also allowing for adaptive complexity \citep{behroozi2019}.
At the same time the number of observables that can be generated with such methods increased including galaxy luminosity \citep[e.g.~][]{rodrigues-puebla2017,moster2018,sommerville2018}, gas \citep[e.g.~][]{popping2015}, metallicity \citep[e.g.~][]{rodrigues-puebla2016} and dust \citep[e.g.~][]{Imara_2018}.

Given their capability to target galaxy formation without biasing the model with baryon physics uncertainties, empirical models complement and help to constrain \textit{ab initio} models.
The power of empirical approaches comes from the possibility to infer the DM density field from observations of the biased luminous component \citep{monaco1999,jasche2019,kitaura2019}.
The mock catalogues obtained can be used to build precise co-variance matrices in preparation for assessing the uncertainties on cosmological parameters estimates, that will be inferred from next generation LSS observational campaigns, such as DESI \citep{desi2013} and Euclid \citep{euclid2018}.
Via the usage of empirical models it is possible to considerably speed up the construction of mock catalogues and are the natural framework for forward modeling of the LSS observable properties \citep[see, e.g.~][]{nuza2014,Leclercq_2015,kitaura2019}.
Furthermore, where \textit{ab initio} models have struggled to obtain tight parameter constraints (e.g., on the mechanism for galaxy quenching), empirical models are capable of revealing possibly new un-expected physics \citep[see, e.g.~][]{Behroozi_2012,Behroozi_2015}.

\textit{Ab initio} approaches are tuned to reproduce the LSS of the Universe at the present time, and therefore their reliability in the high redshift regime has to be proven. 
On the other hand, empirical models are by design particularly suitable for addressing the modelling of the high redshift Universe, but they rely on the availability of high redshift observations of the population to be modelled. 

Our motivation for the original development of the Application Programming Interface (API) we present in this work is to study a particular window in the high redshift Universe.
Specifically, our aim is twofold: $i)$ provide a physically robust and efficient way of modelling galaxy populations in the high redshift Universe from a DM-only N-body simulation; $ii)$ test applications, such as the modelling of the distribution of the first sources that started to shed light on the neutral medium, triggering the process called \emph{Reionization}. We expect that this tool could have further applications, especially in the context of cross-correlation of different tracers and/or diffuse backgrounds.

ScamPy provides a python interface that uses the Sub-halo Clustering and Abundance Matching (SCAM) prescription for ``painting'' galaxies on top of DM-only simulations.
The SCAM algorithm is an extension of the classical HOD for defining the galaxy-halo connection.
This class of methods is widely used in the scientific community but specialised software exists only within larger software packages \citep[e.g. the Halotools package from][which is part of the Astropy library collection]{Hearin2017}.
Our intent is to provide the user with a light and versatile interface able to provide perfomances and extensibility with as little dependence to external software as possible.

We have carefully designed the software to exploit the best features of Python and c++ language. Our intent was not only to achieve high performances of our code but also to make it more accessible, to ease cross-platform installation, and to generally set-up a flexible tool.
Since the API we present has been designed to be easily extensible, in the future we will also be able to evolve our current research towards novel directions.
Furthermore, this effort would hopefully also encourage new users to adopt our tool.
As much as experiments are accurately designed to have the longest life-span possible, we have taken care of designing our software for a long term use.

The API relies on an optimized c++ core implementation of the most computationally expensive sections of the algorithm.
This allows, on the one hand, to exploit the performances of a compiled language.
On the other hand, it overcomes the limit on the usage of multi-threading for shared memory parallelisation, as otherwise imposed by the python standard library.

ScamPy embeds two main functionalities: on the one side, it is designed for handling and building mock-galaxy catalogues, based on an user-defined parameterisation.
On the other side, it provides an extremely efficient implementation of the halo-model, which is used to infer the parameters required by the SCAM algorithm.

We provide a framework for loading a DM halo/sub-halo hierarchy, where the haloes are obtained by means of a friends-of-friends algorithm run on top of cosmological N-body simulations, while the substructures are identified using the SUBFIND algorithm \citep{springel2002}.
Nonetheless, thanks to its extensible design, adapting the API for working with simulations obtained by means of approximated methods, such as COLA \citep{cola2013} or PINOCCHIO \citep{pinocchio2002}, would be straightforward.

Once the ScamPy parameters, which regulate the occupation of structures, have been set, we can easily produce the output mock-catalogue by calling the dedicated functions from the same framework we used for loading the DM halo/sub-halo hierarchy.

This work is organized as follows.
In Section~\ref{sec:scam} we describe the Sub-halo Clustering and Abundance Matching technique.
We describe the main components and algorithms that implement the aforementioned scheme inside our API in Section~\ref{sec:library}.
In Section~\ref{sec:v&v}, we show the results of the several tests we have performed in order to validate the functionalities of our API.
We have tested our instrument in a proof-of-concept application of the target problem: in Section~\ref{sec:ioncase}, we study the effect of individual sources injecting ionizing photons in the neutral inter-galactic medium at high redshift.
Finally, in Section~\ref{sec:conclusions} we provide a summary of this work and anticipate the developments we are planning to pursue.


\section{Sub-halo Clustering \& Abundance Matching}
\label{sec:scam}

Our approach for the definition of the hosted-object/hosting-halo connection is based on the Sub-halo Clustering and Abundance Matching (SCAM) technique \citep{Guo2016}.
With the standard HOD approach, hosted objects are associated to each halo employing a prescription which is based on the total halo mass, or on some other mass proxy  (e.g.\ halo peak velocity, velocity dispersion).
On the other side, the SHAM assumes a monotonic relation between some observed object property (e.g.\ luminosity or stellar mass of a galaxy) and a given halo property (e.g.\ halo mass).
While the first approach is capable of, and extensively used for, reproducing the spatial distribution properties of some target population, the second is the standard for providing plain DM haloes and sub-haloes with observational properties that would otherwise require a full-hydrodynamical treatment of the simulation from which these are extracted.

The SCAM prescription aims to combine both approaches, providing a parameterised model to fit both some observable abundance and the clustering properties of the target population.
The approach is nothing more than applying HOD and SHAM in sequence:
\begin{enumerate}
    \item the occupation functions for central, $N_\text{cen}(M_h)$, and satellite galaxies, $N_\text{sat}(M_h)$, depend on a set of defining parameters which can vary in number depending on the shape used. 
    These functions depend on a proxy of the total mass of the host halo. 
    We sample the space of the defining parameters using a Markov-Chain Monte-Carlo (MCMC) to maximize a likelihood built as the sum of the $\chi^2$ of the two measures we want to fit, namely the two-point angular correlation function at a given redshift, $\omega(\theta, z)$, and the average number of sources at a given redshift, $n_g(z)$:
    \begin{equation}
        \label{eq:loglike00}
        \log\mathcal{L} \equiv -\dfrac{1}{2}\biggl(\chi^2_{\omega(\theta, z)} + \chi^2_{n_g(z)}\biggr)\ \text{,}
    \end{equation}
    The analytic form of both $\omega(\theta, z)$ and $n_g(z)$, depending on the same occupation functions $N_\text{cen}(M_h)$ and $N_\text{sat}(M_h)$, can be obtained with the standard halo model \citep[see ][ for a review]{CooraySheth2002}, which we describe in detail in Section~\ref{sec:halomodel}.
    How these occupation functions are used to select which sub-haloes will host our mock objects is reported in Section~\ref{sec:libpopulator}.
    \item Once we get the host halo/subhalo hierarchy with the abundance and clustering properties we want, as guaranteed by Eq.~\eqref{eq:loglike00}, we can apply our SHAM algorithm to link each mass (or, equivalently, mass-proxy) bin with the corresponding luminosity (or observable property) bin.
\end{enumerate}
When these two steps have been performed, the mock-catalogue is built.

\subsection{The halo model}
\label{sec:halomodel}

The modern formulation of the halo-model theory \citep[see ][for a review]{CooraySheth2002} provides a halo-based description of non-linear gravitational clustering that is widely used in literature to infer the underlying DM statistical properties at both low and high redshift.
The key assumption of this model is that the number of galaxies, $N_g$, in a given dark matter halo only depends on the halo mass, $M_h$.
Specifically, if we assume that $N_g(M_h)$ follows a Poisson distribution with mean proportional to the mass of the halo $M_h$, we can write
\begin{align}
    \label{eq:hm_ng00m1}
    \langle N_g \rangle ( M_h ) &\propto M_h \\
    \label{eq:hm_ng00m2}
    \langle N_g ( N_g - 1 ) \rangle ( M_h ) &\propto M_h^2
\end{align}

From these assumptions it is possible to derive correlations of any order as a sum of the contributions of each possible combination of objects identified in single or in multiple haloes.
To get the models required by Eq.~\eqref{eq:loglike00} we only need the 1-point and the 2-point statistics.
We derive the first as the mean abundance of objects at a given redshift.
The average mass density in haloes at redshift $z$ is given by
\begin{equation}
    \label{eq:hm_rho00}
    \overline{\rho}(z) = \int M_h\; n( M_h, z )\; \dd{M_h}
\end{equation}
where $n( M_h, z )$ is the halo mass function.
With Eq.~\eqref{eq:hm_ng00m1} we can then define the average number of objects at redshift $z$, hosted in haloes with mass $M_\text{min} \le M_h \le M_\text{max}$, as 
\begin{equation}
    \label{eq:hm_ng01}
    n_g(z) \equiv \int_{M_\text{min}}^{M_\text{max}} \langle N_g \rangle (M_h)\; n(M_h, z)\; \dd{M_h}\ \text{.}
\end{equation}

Deriving the 2-point correlation function, $\xi(r, z)$, would require to treat with convolutions, we therefore prefer to obtain it by inverse Fourier-transforming the non-linear power spectrum $P(k,z)$, whose derivation can instead be treated with simple multiplications:
\begin{equation}
    \label{eq:hm_xi00}
    \xi(r, z) = \dfrac{1}{2\pi^2} \int_{k_\text{min}}^{k_\text{max}} \dd{k} k^2 P(k, z) \dfrac{\sin(k r)}{k r}\ \text{.}
\end{equation}
$P(k, z)$ can be expressed as the sum of the contribution of two terms:
\begin{equation}
    \label{eq:hm_pk00}
    P(k, z) = P_\text{1h}(k, z) + P_\text{2h}(k, z)\ \text{,}
\end{equation}
where the first, dubbed \textit{1-halo term}, results from the correlation among objects belonging to the same halo, while the second, dubbed \textit{2-halo term}, gives the correlation between objects belonging to two different haloes.

The 1-halo term in real space is the convolution of two similar profiles of shape
\begin{equation}
    \label{eq:hm_uk00}
    \begin{split}
    \widetilde{u} (k, z|M_h) =& \dfrac{4 \pi \rho_s r_s^3}{M_h}\biggl\{ \sin(k r_s)\bigl[\Si( (1 + c) k r_s ) - \Si(k r_s)\bigr] \\
    &- \dfrac{\sin(c k r_s)}{(1 + c) k r_s} - \cos(k r_s)\bigl[\Ci( (1 + c) k r_s ) - \Ci(k r_s)\bigr]\biggr\}\;\text{,}
    \end{split}
\end{equation}
where $c$ is the halo concentration, $\rho_s$ and $r_s$ are, respectively, the scale density and radius of the NFW profile and the sine and cosine integrals are defined as
\begin{equation}
    \label{eq:hm_sici}
    \Ci(x) = \int_t^\infty \dfrac{\cos t}{t} \dd{t}\ \text{and}\ \Si(x) = \int_0^x \dfrac{\sin t}{t}\dd{t}\;\text{.}
\end{equation}
Eq.~\eqref{eq:hm_uk00} provides the Fourier transform of the dark matter distribution within a halo of mass $M_h$ at redshift $z$.
Weighting this profile by the total number density of pairs, $n(M_h) (M_h/\overline{\rho})^2$, contributed by haloes of mass $M_h$, leads to the expression for the 1-halo term:
\begin{equation}
    \label{eq:hm_pk1h00}
    \begin{split}
        &P_\text{1h}(k, z) \equiv \int n(M_h, z)\, \biggl(\dfrac{M_h}{\overline{\rho}}\biggr)^2 \bigl|\widetilde{u} (k, z|M_h)\bigr|^2 \dd{M_h} =\\
        &= \dfrac{1}{n_g^2(z)} \int_{M_{\text{min}}}^{M_\text{max}} \langle N_{\text{g}}(N_{\text{g}}-1)\rangle(M_h)\; n(M_h, z)\;  \bigl|\widetilde{u}(k, z| M_h)\bigr|^2 \dd{M_h}\text{,}
    \end{split}
\end{equation}
with $n(M_h, z)$ the halo mass function for host haloes of mass $M_h$ at redshift $z$ and where, in the second equivalence, we used Eqs.~\eqref{eq:hm_ng00m2}~and~\eqref{eq:hm_ng01} to substitute the ratio $\bigl(M_h/\overline{\rho}\bigr)^2$.

The derivation of the 2-halo term is more complex and for a complete discussion the reader should refer to \cite{CooraySheth2002}.
Let us just say that, for most of the applications, it is enough to express the power spectrum in its linear form.
Corrections to this approximation are mostly affecting the small-scales which are almost entirely dominated by the 1-halo component.
This is mostly because the 2-halo term depends on the biasing factor which on large scales is deterministic.
We therefore have that, in real-space, the power coming from correlations between objects belonging to two separate haloes is expressed as the product between the convolution of two terms and the biased linear correlation function (i.e. $b_h'(M_h') b_h(M_h'') \xi_\text{lin}(r,z)$).
The two terms in the convolution provide the product between the Fourier-space density profile $\widetilde{u} (k, z|M_h)$, weighted by the total number density of objects within that particular halo (i.e.\ $n(M_h)(M_h/\overline{\rho})$). 
In Fourier space, we therefore have
\begin{equation}
    \label{eq:hm_pk2h00}
    \begin{split}
        P_\text{2h}(k, z) \equiv& \int n( M_h' )\; \dfrac{M_h'}{\overline{\rho}}\; \widetilde{u}( k, z| M_h' )\; b(M_h')\; \dd{M_h'} \\
        & \int n( M_h'' )\; \dfrac{M_h''}{\overline{\rho}}\; \widetilde{u}( k, z| M_h'' )\; b(M_h'')\; P_\text{lin}(k,z)\; \dd{M_h''} = \\
        = \dfrac{P_\text{lin}(k,z)}{n_g^2(z)} & \biggl[ \int_{M_{\text{min}}}^{M_\text{max}} \langle N_\text{g} \rangle(M_h)\; n(M_h)\; b(M_h, z )\;\widetilde{u}_h (k, z | M_h)\;\dd{M_h} \biggr]^2
    \end{split} 
\end{equation}
with $b(M_h)$ the halo bias and $P_\text{lin}(k,z)$ the linear matter power spectrum evolved up to redshift $z$.
For going from the first to the second equivalence we have to make two assumptions.
First we assume self-similarity between haloes. 
This means that the two nested integrals in $\dd{M_h'}$ and $\dd{M_h''}$ are equivalent to the square of the integral in the rightmost expression.
Secondly, we make use of Eqs.~\eqref{eq:hm_ng00m1}~and~\eqref{eq:hm_ng01} to substitute the ratio $M_h/\overline{\rho}$.

The average number of galaxies within a single halo can be decomposed into the sum 
\begin{equation}
    \label{eq:hm_Ng01}
    \langle N_g \rangle(M_h) \equiv N_{\text{cen}} (M_h) + N_{\text{sat}} (M_h)
\end{equation}
where $N_\text{cen}(M_h)$ is the probability to have a central galaxy in a halo of mass $M_h$, while $N_\text{sat}(M_h)$ is the average number of satellite galaxies per halo of mass $M_h$.
These two quantities are precisely the occupation functions we already mentioned in Section~\ref{sec:scam}.
Given that no physics motivated functional form exists for $N_\text{cen}(M_h)$ and $N_\text{sat}(M_h)$, usually, they are parameterised.
By tuning this parameterisation we obtain the prescription for defining the hosted-object/hosting-halo connection.

With the decomposition of Eq.~\eqref{eq:hm_Ng01}, we can approximate Eq.~\eqref{eq:hm_ng00m2} to 
\begin{equation}
\label{eq:hm_NcNs00}
\begin{split}
\langle N_{\text{g}}(N_{\text{g}}-1)\rangle (M_h) &\approx \langle N_{\text{cen}}N_{\text{sat}}\rangle (M_h) + 2\,\langle N_{\text{sat}}(N_{\text{sat}}-1)\rangle (M_h) \approx\\
&\approx N_\text{cen}( M_h )\,N_\text{sat}( M_h ) + N_\text{sat}^2( M_h )
\end{split}
\end{equation}
Thus we can further decompose the 1-halo term of the power spectrum as the combination of power given by {\em central-satellite} couples (cs) and {\em satellite-satellite} couples (ss):
\begin{equation}
\label{eq:hm_Pk1h01}
P_{1h}(k, z) \approx P_{cs}(k, z) + P_{ss}(k, z)\ \text{,}
\end{equation}

When dealing with observations, it is often more useful to derive an expression for the projected correlation function, $\omega(r_p, z)$, where $r_p$ is the projected distance between two objects, assuming flat-sky.
From the Limber approximation \citep{Limber1953} we have
\begin{equation}
    \label{eq:hm_wr00}
    \begin{split}
    \omega (r_p, z) &= \mathcal{A}\bigl[\xi(r, z)\bigr] = \mathcal{A}\bigl\{\mathcal{F}\bigl[P( k, z )\bigr]\bigr\} = \mathcal{H}_0\bigl[P( k, z )\bigr] =\\
    &= \dfrac{1}{2\pi}\int k\; P(k, z)\; J_0( r_p k )\; \dd{k}
    \end{split}
\end{equation}
where $J_0( x )$ is the $0^\text{th}$-order Bessel function of the first kind.
Reading the expression above from left to right, we can get the projected correlation function by Abel-projecting the 3D correlation function $\xi(r,z)$.
From the definition in Eq.~\eqref{eq:hm_xi00}, $\omega(r_p, z)$ is therefore obtained by Abel-transforming the Fourier transform of the power spectrum.
This is equivalent to perform a zeroth-order Hankel transform of the power spectrum, which leads to the last equivalance in Eq.~\eqref{eq:hm_wr00}.

Eq.~\eqref{eq:hm_wr00} though, is valid as long as we are able to measure distances directly in an infinitesimal redshift bin, which is not realistic.
Our projected distance depends on the angular separation, $\theta$, and the cosmological distance, $d_C(z)$, of the observed object
\begin{equation}
    \label{eq:hm_rp00}
    r_p(\theta, z) = \theta \cdot d_C(z)\ \text{.}
\end{equation}
By projecting the objects in our lightcone on a flat surface at the target redshift, we are summing up the contribution of all the objects along the line of sight.
Therefore the two-point angular correlation function can be expressed as
\begin{equation}
    \label{eq:hm_wt00}
        \omega ( \theta, z ) = \int \dfrac{\dd{V}(z)}{\dd{z}} \mathscr{N}^2(z)\; \omega[ r_p(\theta, z), z ] \dd{z}
\end{equation}
where $\frac{\dd{V}(z)}{\dd{z}}$ is the comoving volume unit and $\mathscr{N}(z)$ is the normalized redshift distribution of the target population.
If we assume that $\omega(\theta, z)$ is approximately constant in the redshift interval $[z_1, z_2]$, we can then write 
\begin{equation}
    \label{eq:hm_wt01}
    \omega ( \theta, z ) \approx  \biggl[ \int_{z_1}^{z_2} \dd{z} \dfrac{\dd{V}(z)}{\dd{z}} \mathscr{N}^2(z) \biggr] \cdot \omega[ r_p( \theta ), \overline{z} ) ]
\end{equation}
where $\overline{z}$ is the mean redshift of the objects in the interval.


\section{The ScamPy library}
\label{sec:library}

In this Section we introduce ScamPy, our highly-optimized and flexible API for ``painting'' an observed population on top of the DM-halo/subhalo hierarchy obtained from DM-only N-body simulations.
We will give here a general overview of the algorithm on which our API is based.
We refer the reader to Appendix~\ref{sec:apdxbushido} for a description of the key aspects of our hybrid \textsc{c++}/Python implementation, where we point out how the package is intended for future expansion and further optimization.
We also provide, at the end of this Section, a brief discussion of the API performances.
For a detailed analysis, the reader should refer to Appendix~\ref{sec:apdxperfbench}.
The source code and a guide for installing the library can be obtained by cloning the GitHub repository of the project (\href{https://github.com/TommasoRonconi/scampy}{github.com/TommasoRonconi/scampy}).
The full documentation of ScamPy, with a set of examples and tutorials, is available on the website (\href{https://scampy.readthedocs.io}{scampy.readthedocs.io}) of the package.

\subsection{Algorithm overview}
\label{sec:liboverview}
 
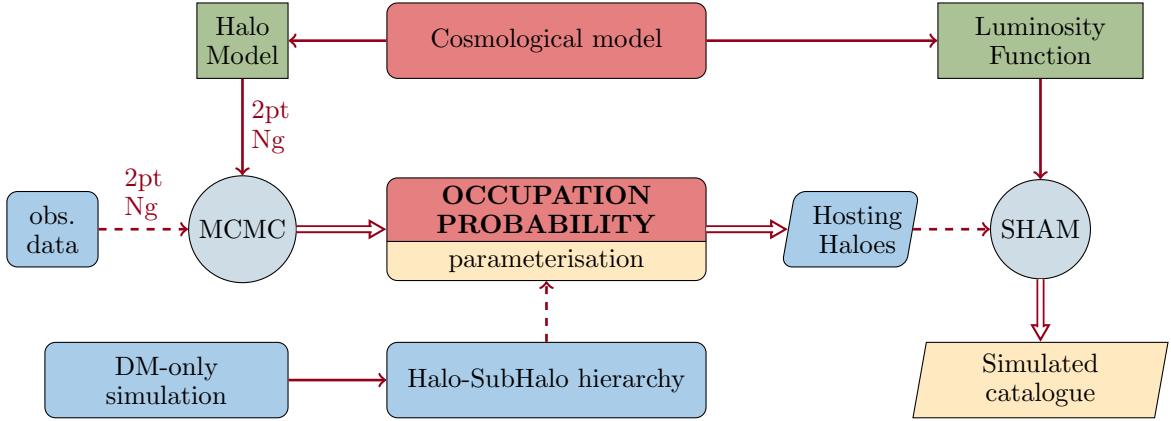
\begin{figure*}
    \centering
    \begin{tikzpicture}[x=1cm, y=10cm, node distance=1cm, font=\large]

    \node (start) [ buildblocksplit, 
                    text width=4cm, 
                    xshift=-2cm ] {\textbf{OCCUPATION PROBABILITY} 
                    \nodepart{second} parameterisation};
    \node (mcmc) [ process, 
                   left of=start, 
                   xshift=-3cm ] {MCMC}; 
    \node (hosts) [ inoutblock, 
                    text width=1cm,
                    right of=start, 
                    xshift=3cm 
                    ] {Hosting\\ Haloes};
    \node (sham) [ process, 
                   right of=hosts, 
                   xshift=1.5cm ] {SHAM};
    \node (gxys) [ outblock, 
                   below of=sham,
                   text width=2.5cm,
                   yshift=-1cm] {Simulated catalogue};
    \node (halomod) [ decision, 
                      text width=1cm,
                      above of=mcmc, 
                      yshift=1.5cm ] {Halo Model};
    \node (cosmo) [ buildblock,
                    above of=start,
                    text width=4cm,
                    yshift=1.5cm ] {Cosmological model};
    \node (lumfunc) [ decision,
                      above of=sham,
                      text width=2.5cm,
                      yshift=1.5cm ] {Luminosity Function};
    \node (halocat) [ inblock, 
                      below of=start,
                      text width=4cm,
                      yshift=-1cm ] {Halo-SubHalo hierarchy};
    \node (dmcat) [ inblock,
                    left of=halocat,
                    text width=3cm,
                    xshift=-4cm ] {DM-only simulation};
    \node (observ) [ inblock,
                     left of=mcmc,
                     xshift=-1.5cm,
                     text width=1cm ] {obs. data};
    \draw [vecArrow] (mcmc) -- (start);
    \draw [innerWhite] (mcmc) -- (start);
    \draw [arrow] (halomod) -- node[right,text width=0.5cm] {2pt\\Ng} (mcmc);
    \draw [dashedarrow] (observ) -- node[above,text width=0.5cm] {2pt\\Ng} (mcmc);
    \draw [arrow] (cosmo) -- (halomod);
    \draw [arrow] (cosmo) -- (lumfunc);
    \draw [dashedarrow] (halocat) -- (start);
    \draw [arrow] (dmcat) -- (halocat); 
    \draw [vecArrow] (start) -- (hosts); 
    \draw [innerWhite] (start) -- (hosts); 
    \draw [dashedarrow] (hosts) -- (sham);
    \draw [arrow] (lumfunc) -- (sham);
    \draw [vecArrow] (sham) -- (gxys);
    \draw [innerWhite] (sham) -- (gxys);

\end{tikzpicture}
    \caption{Flowchart describing the main components of the algorithm. In red the two main kernel modules. Green rectangles dub models from which the workflow depends. Round gray circles are for engines that operate on some inputs. Cyan is for inputs, yellow and parallelograms for outputs.}
    \label{fig:alg_flowchart}
\end{figure*}
In Figure~\ref{fig:alg_flowchart}, we give a schematic view of the main components of the ScamPy package.
All the framework is centred around the occupation probabilities, namely $N_\text{cen}(M_h)$ and $N_\text{sat}(M_h)$, which define the average numbers of, respectively, central and satellite galaxies hosted within each halo. 
Several parameterisations of these two functional forms exist.
One of the most widely used is the standard 5-parameters HOD model \citep{Zheng_2007,Zheng_2009}, with the probability of having a central galaxy given by an activation function and the number distribution of satellite galaxies given by a power-law:
\begin{align}
    \label{eq:alg_hod01}
    N_\text{cen}(M_h) &= \dfrac{1}{2}\biggl[1 + \erf\biggl(\dfrac{\log M - \log M_{\text{min}}}{\sigma_{\log{M_h}}}\biggr)\biggr]\\
    \label{eq:alg_hod02}
    N_\text{sat}(M_h) &= \biggl(\dfrac{M_h - M_\text{cut}}{M_1}\biggr)^{\alpha_{\text{sat}}}
\end{align}
where $M_\text{min}$ is the characteristic minimum mass of halos that
host central galaxies, $\sigma_{\log{M_h}}$ is the width of this transition, $M_\text{cut}$ is the characteristic cut-off scale for hosting satellites, $M_1$ is a normalization factor, and $\alpha_\text{sat}$ is the power-law slope.
Our API provides users with both an implementation of the Eqs.~\eqref{eq:alg_hod01} and \eqref{eq:alg_hod02}, and the possibility to use their own parameterisation by inheriting from a base \texttt{occupation\_p} class.\footnote{The documentation of the library comprehends a tutorial on how to achieve this.}
Given that both the modeling of the observable statistics (Section~\ref{sec:halomodel}) and the HOD method used for populating DM haloes depend on these functions, we implemented an object that can be shared by both these sections of the API.
As outlined in Fig.~\ref{fig:alg_flowchart}, the parameters of the occupation probabilities can be tuned by running an MCMC sampling.
By using a likelihood as the one exposed in Section~\ref{sec:scam}, the halo-model parameterisation that best fits the observed 1- and 2-point statistics of a target population can be inferred.\footnote{In the documentation website we will provide a step-by-step tutorial using \texttt{Emcee} \citep{emcee2013}.}

The chosen cosmological model acts on top of our working pipeline.
Besides providing the user with a set of cosmographic functions for modifying and analysing results on the fly, it plays two significant roles in the API.
On the one hand, it defines the cosmological functions that are used by the halo model, such as the halo-mass function or the DM density profile in Fourier space.
On the other hand, it provides a set of luminosity functions that the user can associate to the populated catalogue through the SHAM procedure.
This approach is not the only one possible, as users are free to define their own observable property distribution and provide it to the function that is responsible for applying the abundance matching algorithm.

Once the HOD parameterisation and the observable-property distribution have been set, it is possible to populate the halo/sub-halo hierarchy of a DM-only catalogue.
\begin{algorithm}
\caption{Schematic outline of the steps required to obtain a mock galaxy catalogue with ScamPy.}
\begin{algorithmic}
\vspace{1mm}
\State{// \texttt{Load Halo/Subhalo hierarchy}}
\State{// \texttt{(e.g. from SUBFIND algorithm)}}
\State{halo\_cat = catalogue( \emph{ chosen from file } )}
\vspace{3mm}
\State{// \texttt{Choose occupation probability function}}
\State{OPF = OPF( \emph{HOD parameters} )}
\vspace{3mm}
\State{// \texttt{Populate haloes}}
\State{gxy\_array = halo\_cat.populate( model = OPF )}
\vspace{3mm}
\State{// \texttt{Associate luminosities}}
\State{gxy\_array = SHAM( gxy\_array, \emph{SHAM parameters} )}
\vspace{3mm}
\end{algorithmic}
\label{algo:populate}
\end{algorithm}
In Alg.~\ref{algo:populate}, we outline the steps required to populate a halo catalogue with mock observables.
We start from a halo/subhalo hierarchy obtained by means of some algorithm (e.g.\ SUBFIND) that have been run on top of a DM-only simulation.
This is loaded into a \texttt{catalogue} structure that manages the hierarchy dividing the haloes in \textit{central} and \textit{satellite} subhaloes.\footnote{For the case of \textsc{subfind} run on top of a \textsc{gadget} snapshot this can be done automatically using the \texttt{catalogue.read\_from\_gadget()} function. We plan to add similar functions for different halo-finders (e.g.\ \textsc{rockstar}, \cite{rockstar2013}, and \textsc{sparta}, \cite{Diemer_2017}) in future extensions of the library.}

Our \texttt{catalogue} class comes with a \texttt{populate()} member function that takes an object of type occupation probability as argument and returns a trimmed version of the original catalogue in which only the central and satellite haloes hosting an object of the target population are left.
We give a detailed description of this algorithm in Section~\ref{sec:libpopulator}.
When this catalogue is ready, the SHAM algorithm can be run on top of it to associate at each mass a mock-observable property.
Cumulative distributions are monotonic by construction.
Therefore it is quite easy to define a bijective relation between the cumulative mass distribution of haloes and the cumulative observable property distribution of the target population. 
This algorithm is described in Section~\ref{sec:libsham}.

\subsubsection{Populating algorithm}
\label{sec:libpopulator}

Input subhalo catalogues are trimmed into hosting subhalo catalogues by passing to the \texttt{populate()} member function of the class \texttt{catalogue} an object of type \texttt{occupation\_p}.

In Algorithm~\ref{algo:hod}, we describe this halo occupation routine.
\begin{algorithm}[htbp]
    \caption{Description of the \texttt{populate} (model = OPF) function. This is an implementation of the HOD prescription, where the assumptions made to define the halo model (i.e.\ the average number of objects within a halo follows a Poisson distribution with mean $\langle N_g\rangle(M_h)$) are accounted for.}
    \begin{algorithmic}
        \vspace{1mm}
        \State{// \texttt{Iterate over all the haloes in catalogue}}
        \For{ halo in catalogue }
            \vspace{3mm}
            \State{// \texttt{Compute probability of central}}
            \State{$p_\text{cen} \leftarrow \text{model.}N_\text{cen}( $ halo.mass $ )$ }
            \vspace{3mm}
            \State{// \texttt{Define a binomial random variable}}
            \State{select $\leftarrow$ random.Binomial(1, $p_\text{cen}$)}
            \If{ select }
                \State{halo $\leftarrow$ central}
            \EndIf
            \vspace{3mm}
            \State{// \texttt{Compute average number of satellites}}
            \State{$\overline{N}_\text{sat} \leftarrow \text{model.}N_\text{sat}( $ halo.mass $ )$}
            \vspace{3mm}
            \State{// \texttt{Define a Poisson random variable}}
            \State{$N_\text{sat} = $ random.Poisson( $\overline{N}_\text{sat}$ )}
            \State{halo $\leftarrow$ select randomly $N_\text{sat}$ objects among satellites}
        \EndFor
        \vspace{3mm}
    \end{algorithmic}
    \label{algo:hod}
\end{algorithm}
For each halo $i$ in the catalogue, we compute the values of $\langle N_\text{cen}\rangle(M_i)$ and $\langle N_\text{sat}\rangle(M_i)$.
To account for the assumptions made in our derivation of the halo model, we select the number of objects each halo will host by extracting a random number from a Poisson distribution.
For the occupation of the central halo this reduces to extracting a random variable from a Binomial distribution: $N_\text{cen} = \mathscr{B}(1, \langle N_\text{cen}\rangle_i)$.
While, in the case of satellite subhaloes, we extract a random Poisson variable $N_\text{sat} = \mathscr{P}(\langle N_\text{sat}\rangle_i)$, then we randomly select $N_\text{sat}$ satellite subhaloes from those residing in the $i^\text{th}$ halo.
    
\begin{figure}
    \centering
    \includegraphics[width=0.45\textwidth]{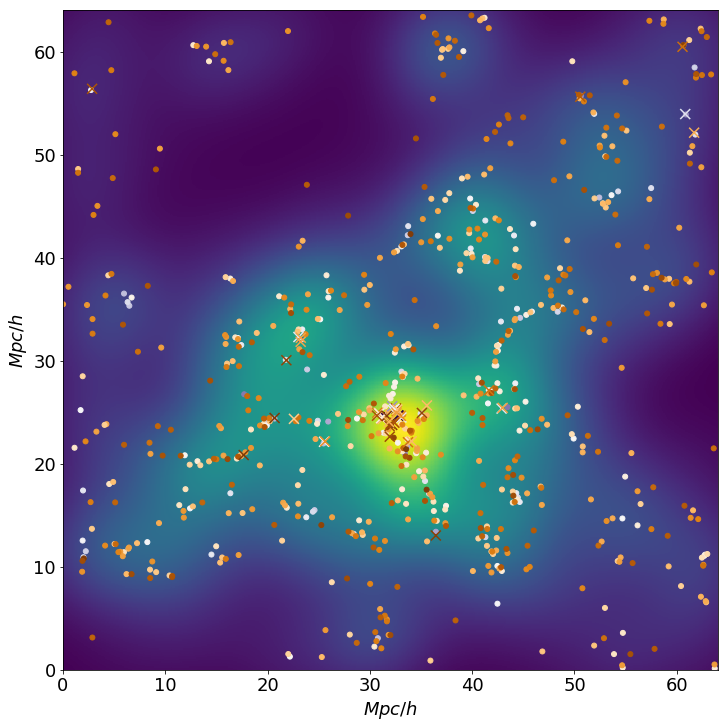}
    \caption{A $4\ \text{Mpc}/h$ thick slice of a populated catalogue obtained from a DM-only simulation with $64\ \text{Mpc}/h$ box side lenght. The colour code on the background shows the smoothed DM density field (with density increasing going from darker to brighter regions) while the markers show our mock galaxies (with color representing lower to higher luminosity going from brighter to darker). Circles are for centrals and crosses for satellites.}
    \label{fig:pop_cat_slice}
\end{figure}

In Fig.~\ref{fig:pop_cat_slice}, we show a $4\ \text{Mpc}/h$ thick slice of a simulation with box side lenght of $64\ \text{Mpc}/h$, the background colour code represents the density field traced by all the subhaloes found by the SUBFIND algorithm, smoothed with a Gaussian filter, while the markers show the positions of the subhaloes selected by the populating algorithm. 
We will show in Section~\ref{sec:val1-2pt} that this distribution of objects reproduces the observed statistics.
It is possible to notice how the markers trace the spatial distribution of the underlying DM density field.

\subsubsection{Abundance matching algorithm}
\label{sec:libsham}

When the host subhaloes have been selected we can run the last step of our algorithm.
The \texttt{abundance\_matching()} function implements the SHAM prescription to associate to each subhalo an observable property (e.g.\ a luminosity or the star formation rate of a galaxy).
This is achieved by defining a bijective relation between the cumulative distribution of subhaloes as a function of their mass and the cumulative distribution of the property we want to associate them.

\begin{figure*}
    \centering
    \includegraphics[width=\textwidth]{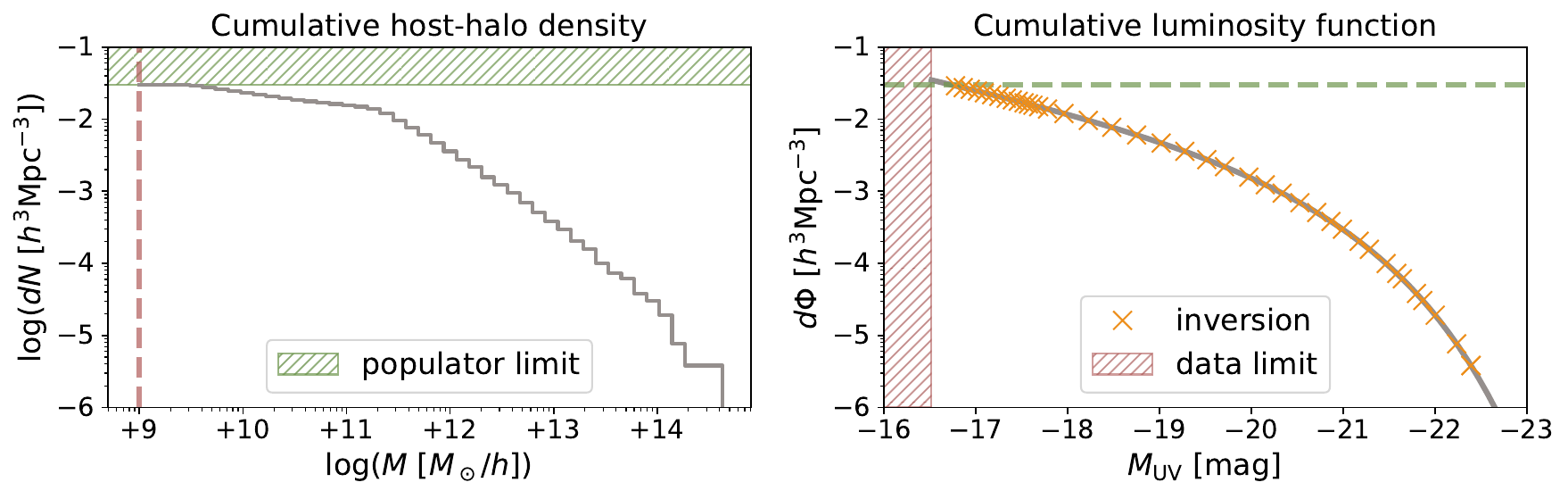}
    \caption{Abundance matching scheme. \textit{Left panel:} cumulative number density of host subhaloes divided into regular logarithmic bins (solid gray step-line). The green dashed band shows the limit imposed by the resolution in mass of the simulation, which is inherited by the host catalogue obtained with the populator algorithm. \textit{Right panel:} cumulative luminosity function (solid gray line). The dashed red band shows the limit imposed by the magnitude limit of the observed target population. Orange crosses mark the positions, bin-per-bin, of the abundances measured in each bin of the left panel. In both the left and the right panel we reported with a dashed line of corresponding colour the limit imposed by the resolution of the catalogue (dashed green line in right panel) and by the magnitude limit of the survey (dashed red line in the left panel).}
    \label{fig:sham}
\end{figure*}

An example of this procedure is shown in Fig.~\ref{fig:sham}.
We want to set, for each subhalo, the UV luminosity of the galaxy it hosts.
In the left panel, we show the cumulative mass-distribution of subhaloes, $\dd{N}(M_\text{subhalo})$, with the dashed green region being the mass resolution limit of the DM subhaloes in our simulation after the populating algorithm has been applied.
On the right panel we show the cumulative UV luminosity function, which is given by the integral
\begin{equation}
    \label{eq:sham_Phi00}
    \Phi(M^\text{UV} < M_\text{lim}^\text{UV}) = \int_{-\infty}^{M_\text{lim}^\text{UV}} \dfrac{\dd{\Phi}}{\dd{M^\text{UV}}}\dd{M^\text{UV}}
\end{equation}
where $M_\text{lim}^\text{UV}$ is the limiting magnitude of the survey data we want to reproduce (marked by a dashed red region in Fig~\ref{fig:sham}).
We find the abundance corresponding to each mass bin (gray step line in the left panel) and we compute the corresponding luminosity by inverting the cumulative luminosity function obtained with Eq.~\eqref{eq:sham_Phi00}:
\begin{equation}
\label{eq:sham_Phi01}
M^\text{UV}(M_\text{subhalo}) = \Phi^{-1}[\dd{N}(M_\text{subhalo})]\ \text{.}
\end{equation}
The result of this matching is shown by the orange crosses in the right panel of Fig.~\ref{fig:sham}.
At the time we are writing, the scatter around the distribution of luminosities can be controlled by tuning the bin-width used to measure $\dd{N}(M_\text{subhalo})$.
We plan to extend this functionality of the API by adding a parameter for tuning this scatter to the value chosen by the user. 

In Figure~\ref{fig:pop_cat_slice}, the colour gradient of markers highlights the increase in their associated luminosity (from lighter to darker colour, going from fainter to brighter object). 

\subsection{Performance discussion}
\label{sec:perfparal}

In this Section we comment some design choices made for optimizing the performances of the library.
A detailed discussion on the implementation is provided in Appendix~\ref{sec:apdxbushido}.
For some benchmarking measures of the API performances we refer the reader to Appendix~\ref{sec:apdxperfbench}.

With our hybrid implementation we have deployed a library that exploits both the performance efficiency of a compiled language (\textsc{c++}) as well as the flexibility of an interpreted language (Python). 
The \textsc{c++} core of the library has multi-threaded sections to run in parallel the most computationally demanding calculations of the workflow. 
This choice has been made in order to circumvent the Global Interpreter Lock (GIL) which  would otherwise force the Python application to run on a single thread.
Spawning threads directly from the \textsc{c++} core is more efficient than using most of the Python packages for multi-threading. 
Nonetheless this behaviour can be suppressed by setting the corresponding environment variable accordingly.

Functions that do not spawn threads by default are those meant for modelling cosmological statistics (e.g. clustering and abundances) and functions that have a pure Python implementation.

Since the former functions would ideally be used in a MCMC framework, they run on a single thread. 
In this way all the cores of the CPU are available for parallelizing the parameter space sampling.
Nevertheless, we have carefully optimized and benchmarked such functions (more details in Appendix~\ref{sec:apdxbushido} and~\ref{sec:apdxperfbench}, respectively):
on a single thread, one single computation of the likelihood in Eq.~\eqref{eq:loglike00} for a typical problem size (i.e. a dataset with approximately $10$ degrees of freedom) takes approximately $1$ millisecond, running on a common laptop.

Concerning the pure Python implementation, the \texttt{catalogue} class and the \texttt{populate} function are the only sections of the library that would gain from a multi-threaded core, given that they operate on lists of independent objects. 
Such functionality has not been implemented yet but would be a natural evolution of our library.
At the current state of the implementation, both these operations take times in the order of $1\div10$ seconds for a sub-halo table with $10^{4\div5}$ objects.
Loading and populating sub-haloes have $\mathcal{O}(N_\text{sub})$ scaling, where $N_\text{sub}$ is the number of sub-haloes loaded into the catalogue.
The dependence of these time measurements to the machine architecture is negligible.

\section{Verification \& Validation}
\label{sec:v&v}

We have extensively tested all the functions building up our API in all their unitary components.
We developed a testing machinery, included in the official repository of the project, to run these tests in a continuous integration environment.
This will both guarantee consistency during future expansions of the library, as long as providing users with a quick check that the build have been completed successfully.

In this Section, we show that our machinery is producing the expected results.
Specifically, in Section \ref{sec:val1-2pt} we show that the mock-catalogues obtained with ScamPy reproduce the observables we want.
We have also tested our API for the accuracy in reproducing cross-correlations in Section \ref{sec:valcross}.
Even though there is no instruction in the algorithm that guarantees this behaviour, using the halo model it is trivial to obtain predictions for the cross-correlation of two different populations of objects.

\begin{table}
    \centering
    \caption{Fiducial values of the HOD parameters at the different redshifts inspected. The HOD model is defined by Eqs.~\eqref{eq:alg_hod01}~and~\eqref{eq:alg_hod02}.}
    \label{tab:hod_fid}
    \begin{tabular}{cccccc}
\toprule
$z$ & $M_\text{min}$ & $\sigma_{\log M}$ & $M_\text{cut}$ & $M_1$ & $\alpha_\text{sat}$ \\
& $[10^{10}\ M_\odot/h]$ && $[M_\odot/h]$ & $[10^{12}\ M_\odot/h]$&\\
\midrule
$0$ & $5.0$ & $0.3$ & $0.0$ & $1.0$ & $1.0$ \\
$2$ & $2.0$ & $0.3$ & $0.0$ & $1.0$ & $1.0$ \\
$4$ & $2.0$ & $0.3$ & $0.0$ & $1.0$ & $1.0$ \\
$6$ & $1.0$ & $0.3$ & $0.0$ & $0.5$ & $1.2$ \\
$8$ & $1.5$ & $0.2$ & $0.0$ & $0.5$ & $1.2$ \\
\noalign{\smallskip}\hline \bottomrule
\end{tabular}
\end{table}
All the validation tests have been obtained by assuming a set of reasonable values for the HOD parameters of Eqs.~\eqref{eq:alg_hod01}~and~\eqref{eq:alg_hod02}.
All the parameters used are listed in Table~\ref{tab:hod_fid}.
We will refer to these sets of parameters as \textit{fiducial model} in the rest of this manuscript.

The resulting occupation probabilities have been then used to populate a set of halo/subhalo catalogues.
These catalogues have been obtained by running on the fly the FoF and SUBFIND \citep{springel2002} algorithms on top of a set of cosmological N-body simulations.
The DM snapshots have been obtained by running the (non-public) \textsc{P-GADGET-3} N-body code \citep[which is derived from the \textsc{GADGET-2} code,][]{gadget}.
In Table~\ref{tab:cosmo_sim} we list the different simulation boxes we used for testing the library.
Given the large computational cost of running high resolution N-body codes, only the \texttt{lowres} simulation box has been evolved up to redshift $z = 0$, while we stopped the others at redshift $z = 2$.
\begin{table}
    \centering
    \caption{Our set of cosmological simulations with the corresponding relevant physical quantities: $N_\text{part}$ is the total number of DM particles; $M_\text{part}$ is the mass of each particle; $L_\text{box-side}$ is the side lenght of the simulation box; $z_\text{min}$ is the minimum redshift up to which the simulation has been evolved.}
    \label{tab:cosmo_sim}
    \begin{tabular}{cccccc}
\toprule
name & $N_\text{part}$ & $M_\text{part}$ & $L_\text{box-side}$ & $z_\text{min}$\\ \midrule
\texttt{lowres} & $512^3$   & $8.13\times10^7\ M_\odot/h$ & $64\ \text{Mpc}/h$ & $0$ \\
\texttt{midres} & $1024^3$  & $1.02\times10^7\ M_\odot/h$ & $64\ \text{Mpc}/h$ & $2$ \\
\texttt{highres} & $1024^3$ & $1.27\times10^6\ M_\odot/h$ & $32\ \text{Mpc}/h$ & $2$ \\ \noalign{\smallskip}\hline \bottomrule
\end{tabular}
\end{table}
The cosmological parameters used for all these simulations are summarised in Table~\ref{tab:cosmo_params}.
\begin{table}
    \centering
    \caption{Fiducial cosmological parameters of the N-body simulations used in this work.}
    \label{tab:cosmo_params}
    \begin{tabular}{cccccc}
\toprule
$h$ & $\Omega_\text{CDM}$ & $\Omega_b$ & $\Omega_\Lambda$ & $\sigma_8$ & $n_s$ \\ \midrule
$0.7$ & $0.3$ & $0.045$ & $0.7$ & $0.8$ & $0.96$ \\\noalign{\smallskip}\hline \bottomrule
\end{tabular}
\end{table}

\subsection{Observables}
\label{sec:val1-2pt}

Here we present measurements obtained after both the populating algorithm of Section~\ref{sec:libpopulator} and the abundance matching algorithm of Section~\ref{sec:libsham} have been applied to the DM-only input catalogue.
Applying the abundance matching algorithm does not modify the content of the populated catalogue, besides associating to each mass an additional observable-property.

\begin{figure*}
    \centering
    \includegraphics[width=0.95\textwidth]{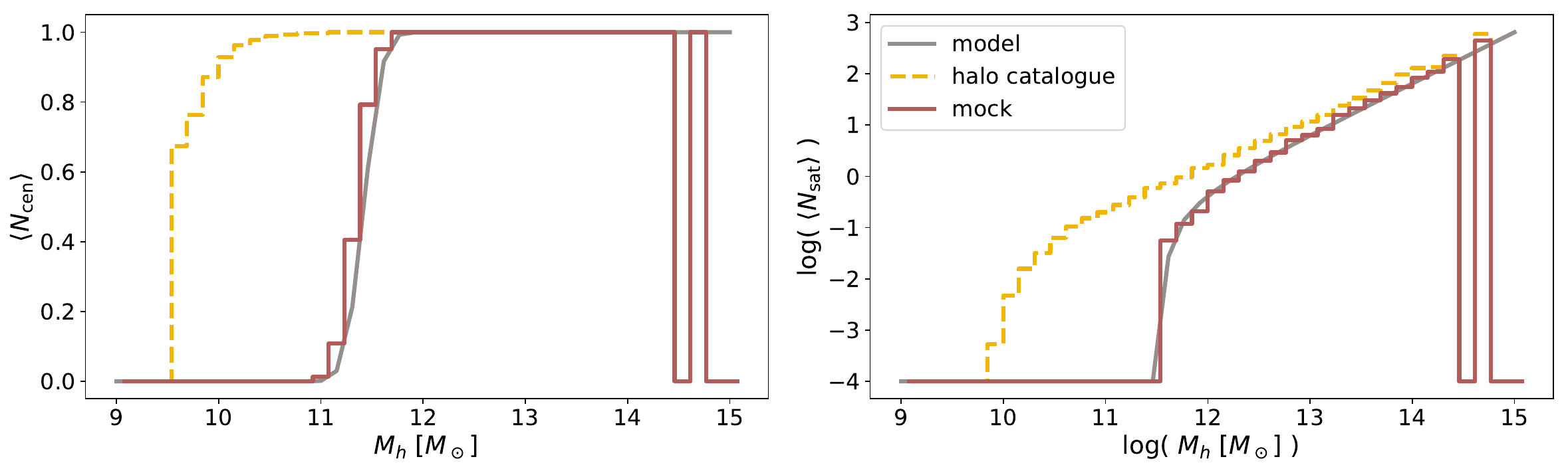}
    \caption{Occupation probability functions for central subhaloes (\emph{left~panel}) and satellite subhaloes (\emph{right~panel}). The gray solid line marks the $z=0$ fiducial model of Table~\ref{tab:hod_fid}. The yellow step-wise dashed line and the red step-wise solid line mark the distributions measured on the subhalo catalogue before and after having applied our populating algorithm.}
    \label{fig:ocp_probs}
\end{figure*}
In the two panels of Fig. \ref{fig:ocp_probs} we show the abundances of central and satellite subhaloes, as a function of the halo mass.
The dashed yellow step-lines show the distribution in the DM-only input catalogue, while the gray solid line marks the distribution defined by the fiducial occupation probability functions.
By applying the populating algorithm (Section \ref{sec:libpopulator}) to the input catalogue we obtain the distributions marked by the solid red step-lines, which are in perfect agreement with the expected distribution.

We then draw a random Gaussian sample around the halo model estimate for the objects abundance and their clustering using the above selection of occupation probabilities (Eqs.~\eqref{eq:hm_ng01} and \eqref{eq:hm_xi00} respectively).
These random samples build up our \emph{mock dataset}.
We then run an MCMC sampling of the parameter space, with the likelihood of Eq.~\eqref{eq:loglike00}, to infer the set of parameters that best fit the mock dataset.
For sampling the parameter space we use the Emcee \citep{emcee2013} Affine Invariant MCMC Ensemble sampler, along with the ScamPy python interface to the halo model estimates of $n_g(z)$ and $\xi(r, z)$.

After having obtained the best-fit parameters, we produce 10 runs of the full pipeline described in Alg.~\ref{algo:populate}.
In doing this, we are producing 10 different realisations of the resulting mock catalogue.
Since the selection of the host-subhaloes is not deterministic, this procedure allows to obtain an estimate of the errors resulting from the assumptions of the halo model.
Finally, we use the Landy-Szalay \citep{LandySzalay1993} estimator in each of the populated catalogues to measure the 2-point correlation function:  
\begin{equation}
    \label{eq:LS_est}
    \xi(r) = \dfrac{DD(r) - 2 DR(r) + RR(r)}{RR(r)}
\end{equation}
where $DD(r)$ is the normalised number of unique pairs of subhaloes with separation $r$, $DR(r)$ is the normalised number of unique pairs between the populated catalogue and a mock sample of objects with random positions, and $RR(r)$ is the normalised number of unique pairs in the
random objects catalogue.
We then measure, with Eq.~\eqref{eq:LS_est}, the clustering in each of the 10 realisations and we compute the mean and standard deviation of these measurements in each radii bin. 

\begin{figure}
    \includegraphics[width=0.45\textwidth]{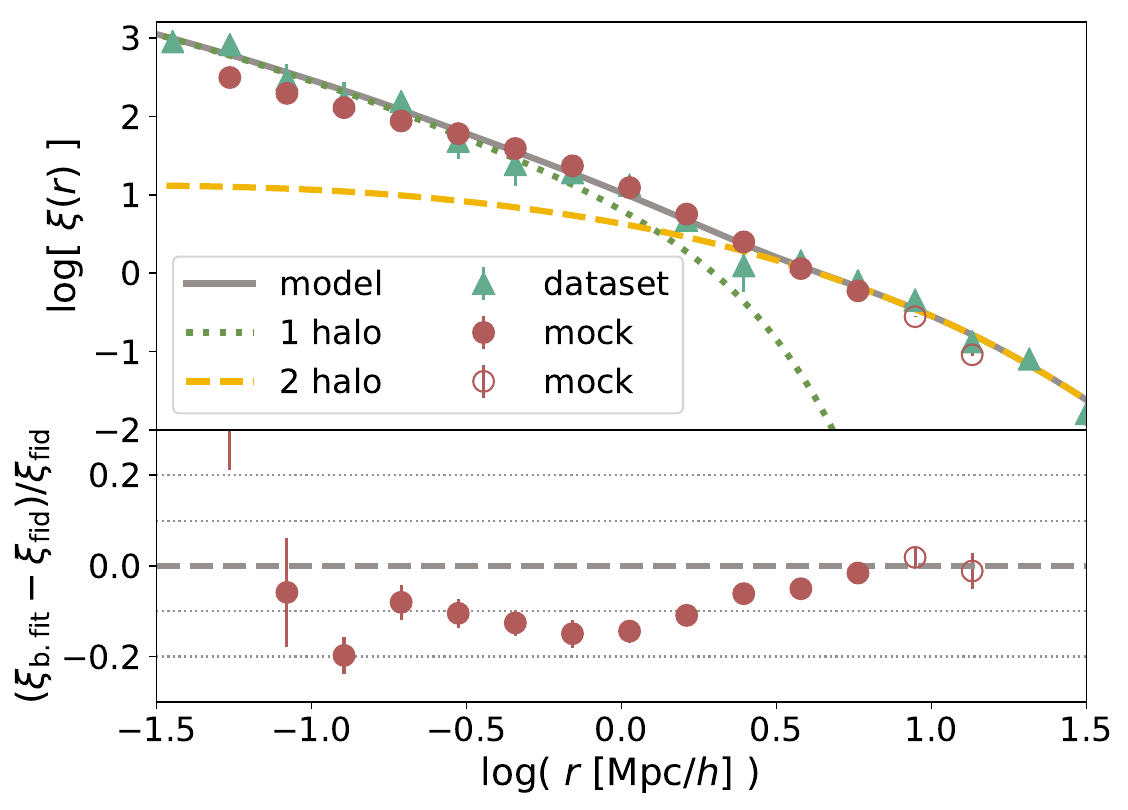}
    \caption{Validation of the two point correlation function at redshift $z=0$. 
    \emph{Upper panel:} comparison between the mock clustering dataset (triangles), the best-fit halo-model prediction (solid line) and the mean and standard deviation of the clustering measured with the Landy-Szalay estimator on the 10 realisations (circles and errors), we also show  the modelled 1-halo (dotted line) and 2-halo (dashed line) terms for reference. \emph{Lower panel:} red markers show the distance ratio  between the measurement obtained on the mock catalogue with the best-fit parameters and the averaged measurement obtained on the mock catalogue with fiducial parameters. The dashed line shows $0\%$ distance between the two.
    In both panels, empty circles mark measurements at $r > 6.4\ \text{Mpc}/h$, where the limited box size affects the precision of the result.}
    \label{fig:2pt_00}
\end{figure}

\begin{figure*}
    \includegraphics[width=0.95\textwidth]{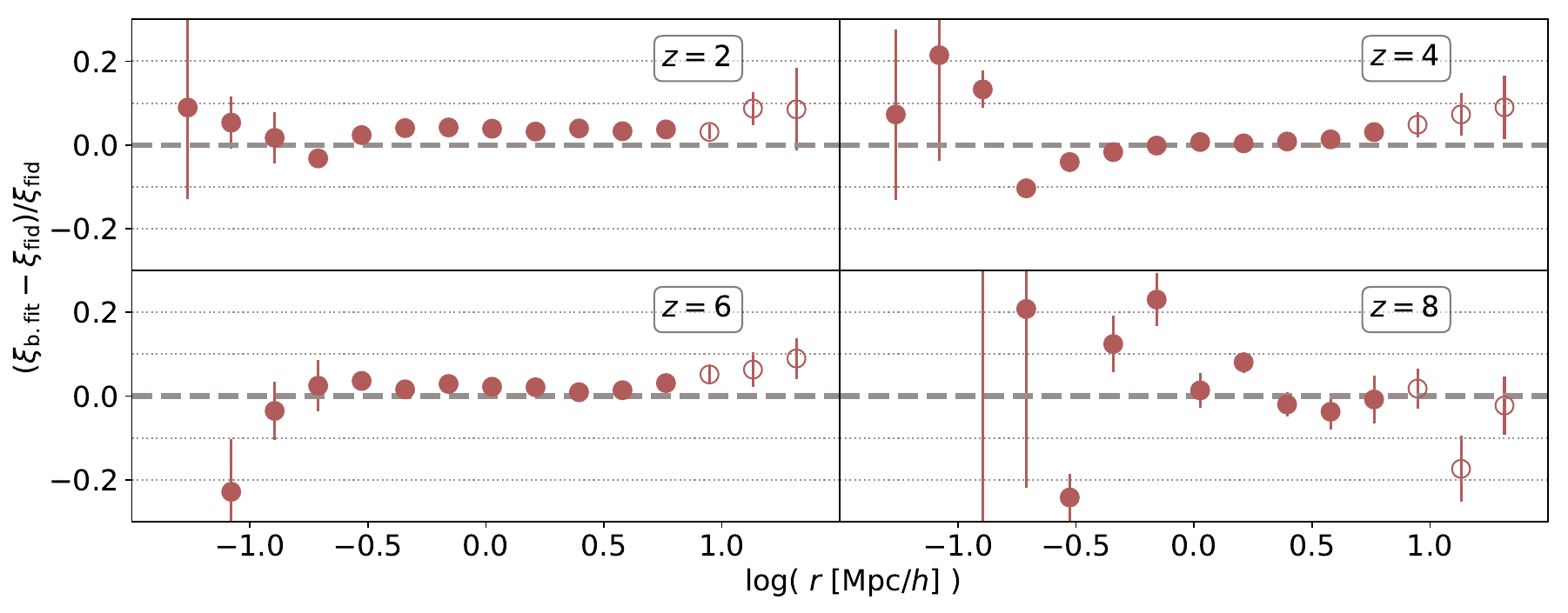}
    \caption{Validation of the two point correlation function at redshift $z=2$ (upper left panel), $z=4$ (upper right panel), $z=6$ (lower left panel), $z=8$ (lower right panel). Red markers show the distance ratio  between the averaged-measurement obtained on the mock catalogue with the best-fit parameters and the measurement obtained on the mock catalogue with fiducial parameters. The dashed line shows $0\%$ distance between the two. As in Fig.~\ref{fig:2pt_00}, empty circles mark measurements at $r > 6.4\ \text{Mpc}/h$}
    \label{fig:2pt_01}
\end{figure*}

The results are shown in Fig.~\ref{fig:2pt_00}, for redshift $z = 0$, and in Fig.~\ref{fig:2pt_01}, for redshifts $z = 2, 4, 6, 8$.
In the upper panel of Fig. \ref{fig:2pt_00} we show the mock dataset with triangle markers and errors, the lines show the halo model best-fit estimate of the 2 point correlation function (with the different contributes of the 1- and 2-halo terms).
The circle markers show the average measure obtained from our set of mock-catalogues.

In the lower panel of Fig. \ref{fig:2pt_00} and in the four panels of Fig. \ref{fig:2pt_01} we show the relative distance between the measure performed on the catalogue populated with the best-fit parameterisation of the occupation probabilities ($\xi_\text{b.fit}$) and on a catalogue populated with the fiducial value of the parameters ($\xi_\text{fid}$).
Comparing the measurements in the two different populated catalogues, instead of comparing with the model itself, guarantees that discrepancies due to box-size and resolution of the simulation used are mitigated in the distance ratio plot.

As it is shown in the lower panel of Fig.~\ref{fig:2pt_00}, at redshift $z=0$, the catalogue populated with the best-fit parameters reproduces the clustering properties of the fiducial catalogue with a distance lower than $15\%$ on most of the scales inspected.

The discrepancies at small scales could depend on several effects.
In particular:
\begin{itemize}
    \item Implementation of the populating algorithm: galaxies can be assigned to satellite haloes randomly or by rank-ordering them based on some property of the sub-halo.
    The two methods might produce different levels of clustering within the halo.
    \item The sub-halo finder used: SUBFIND is known for not resolving completely the sub-halo hierarchy closer to the halo centre.
    \item Simulation resolution: discretization of the dark matter distribution limits the smallest sub-halo that can be modeled.
    This also results in a larger scatter when going at higher redshift (as shown in Fig.~\ref{fig:2pt_01}) where the average size of a halo gets smaller.
\end{itemize}
Concerning the latter point, it is worth to mention that, even thought the measurement is less precise, the average distance from the expected result is still lower than $10\div15\%$.
Finally, in both Fig.~\ref{fig:2pt_00} and Fig.~\ref{fig:2pt_01}, we mark with empty circles measurements taken at radii $r>6.4\ \text{Mpc}/h$, where the limited size of the simulation box affects the statistics.

All the discrepancies we find are a known weakness of the HOD method.
In literature there have been a lot of effort in quantifying and correcting this effect \citep[see e.g.][for two recent works]{Beltz-Mohrmann2019, Hadzhiyska2019}, which, as already mentioned, is thought to result from a concurrence of box-size effects, cosmic-variance and assembly bias.
\cite{Hadzhiyska2019}, in particular, find an average distance of $15\%$ between the HOD prediction and the clustering measured in hydro-dynamical N-body simulations. 

The requirement of reproducing the 1-point statistics of the original catalogue is necessary to have the expected observational property distribution in the output mock-catalogue.
This requirement guarantees that the abundance matching scheme will start associating the observational property from the right position in the cumulative distribution, i.e.\ from the abundance corresponding to the limiting value that said property has in the survey.

\begin{figure}
    \includegraphics[width=0.45\textwidth]{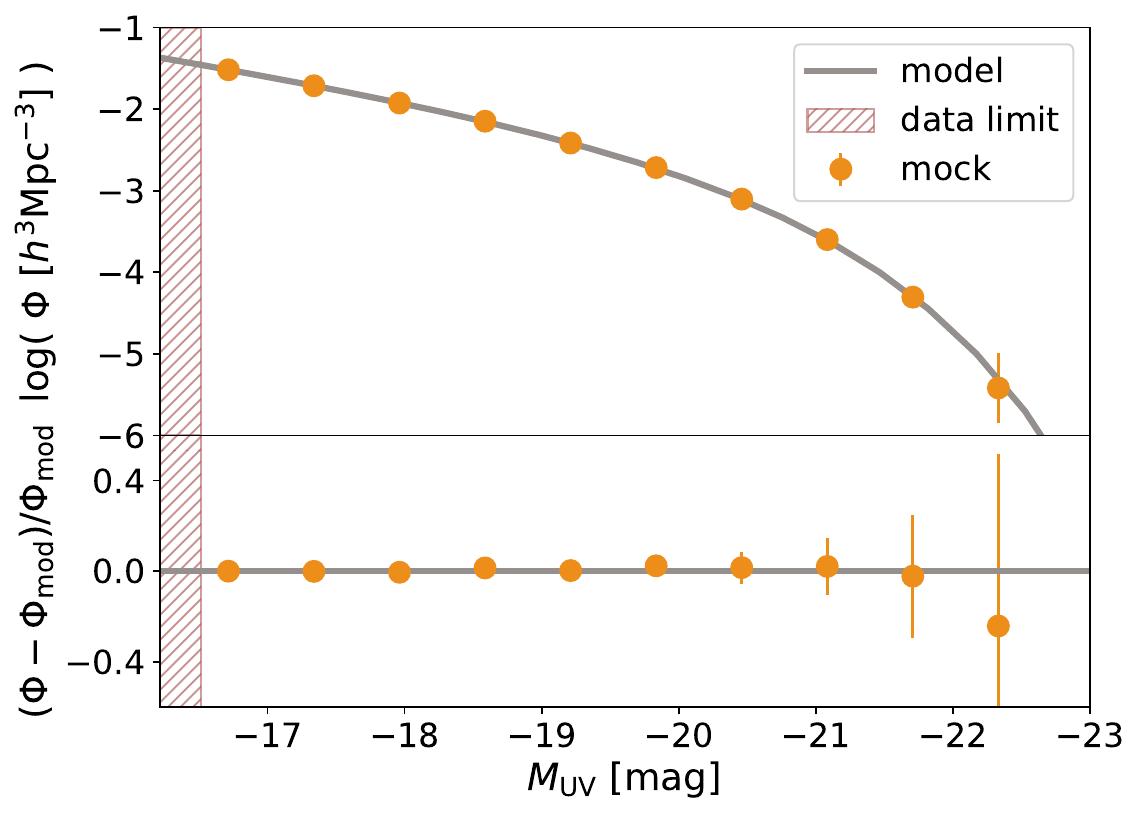}
    \caption{Cumulative luminosity function at redshift $z = 0$. \emph{Upper~panel:} the gray solid line marks the model prediction while the orange circles with errors mark the distribution measured on the populated catalogue. The hatched red region marks the limiting magnitude $M_\text{lim}^\text{UV}$. \emph{Lower~panel:} distance ratio between the luminosity function measured on the populated catalogue and the model.}
    \label{fig:lum_cum}
\end{figure}

In Fig.~\ref{fig:lum_cum}, we show the example case of the UV luminosity function. 
We mark with orange circles the cumulative luminosity function measured on the mock-catalogue after the application of our API.
For comparison we also show the luminosity function model we are matching (gray solid line) and the observation limit of the target population (red dashed region).

As it is shown in the lower panel of Fig.~\ref{fig:lum_cum}, the distance ratio between the expected distribution and the mock distribution is lower than $\approx 10\%$ over all the range of magnitudes.

The halo-model prediction for the total abundance of sources is $n_g^\text{hm}(z) = 3.49\cdot10^{-2}\ [h^3 \text{Mpc}^{-3}]$, while we measure $n_g^\text{pop}(z) = ( 3.06 \pm 0.04 )\cdot10^{-2}\ [h^3 \text{Mpc}^{-3}]$ in the populated catalogue.
Such a miss-match is somehow expected.
In fact, it has been shown \citep[e.g.~][]{sinha2018} that the HOD model presents difficulties when fitting multiple statistics.
For the distribution of sources shown in Fig.~\ref{fig:lum_cum}, we correct this error by extrapolating the limiting magnitude (marked by the hatched region) to the value matching the abundance of the populated galaxies.

\cite{sinha2018} also show that the clustering statistics with the higher constraining power depends on the galaxy population that has to be modeled.
Since we are running our analysis on a simulated dataset, with the aim of validating the computational framework, we are not pushing this analysis further. 
Nevertheless, in a real-life application, the likelihood of Eq.~\eqref{eq:loglike00} should be adjusted considering these observations, depending on the scientific goal of the application. 
The \texttt{halo\_model} class of our library provides a wide range of cosmological statistics that can be modeled, leaving to the users the responsibility of choosing the one that better suits their needs.

\subsection{Multiple populations cross-correlation}
\label{sec:valcross}

Even though in our API there is no prescription for this purpose, it is interesting to test how the framework performs in predicting the cross-correlation between two different populations.
This quantity measures the fractional excess probability, relative to a random distribution, of finding a mock-source of population 1 and a mock-source of population 2, respectively, within infinitesimal volumes separated by a given distance.

It is simple to modify Eqs.~\eqref{eq:hm_pk1h00} and \eqref{eq:hm_pk2h00} to get the expected power spectrum of the cross-correlation \citep{CooraySheth2002}.
For the 1-halo term this is achieved by splitting the $(M_h/\overline{\rho})^2$ of Eq.~\eqref{eq:hm_pk1h00} in the contribution of the two different populations, which leads to the following equation:
\begin{equation}
    \label{eq:hm_cross_1h}
    \begin{split}
    P_\text{1h}^{(1,2)}(k, z) =& \dfrac{1}{n_g^{(1)}(z)n_g^{(2)}(z)}\cdot\\
    &\int_{M_{\text{min}}}^{M_\text{max}} N_g^{(1)}(M_h) N_g^{(2)}(M_h) n_h(M_h) |\widetilde{u}_h (k, M_h, z)|^2\text{d}M_h
    \end{split}
\end{equation}
where quantities referring to the two different populations are marked with the superscripts $(1)$ and $(2)$.

For the case of the 2-halo term, obtaining an expression for the cross-correlation requires to divide the two integrals of Eq.~\eqref{eq:hm_pk2h00} in the contributions of the the two different populations, leading to
\begin{equation}
    \label{eq:hm_cross_2h}
    \begin{split}
        P_\text{2h}^{(1,2)}(k, z) =& \dfrac{P_m(k, z)}{n_g^{(1)}(z)n_g^{(2)}(z)}\ \cdot \\
        & \biggl[\int_{M_{\text{min}}}^{M_\text{max}} N_g^{(1)}(M_h) n_h(M_h) b_h(M_h, z )\widetilde{u}_h (k, M_h, z)\text{d}M_h\biggr]\ \cdot\\
        & \biggl[\int_{M_{\text{min}}}^{M_\text{max}} N_g^{(2)}(M_h) n_h(M_h) b_h(M_h, z )\widetilde{u}_h (k, M_h, z)\text{d}M_h\biggr]
    \end{split}
\end{equation}

We get the cross-correlation of the two mock-populations using a modification \citep{Gonzalez2017} of the Landy-Szalay estimator
\begin{equation}
    \label{eq:cross_estimator}
    \xi^{(1,2)} (r) = \dfrac{D_1 D_2(r) - D_1 R_2(r) - D_2 R_1(r) + R_1 R_2(r)}{R_1 R_2(r)}
\end{equation}
where $D_1 D_2(r)$,  $D_1 R_2(r)$,  $D_2 R_1(r)$ and  $R_1 R_2(r)$ are the normalized data1-data2, data1-random2, data2-random1 and random1-random2 pair counts for a given distance $r$.

\begin{figure}
    \includegraphics[width=0.45\textwidth]{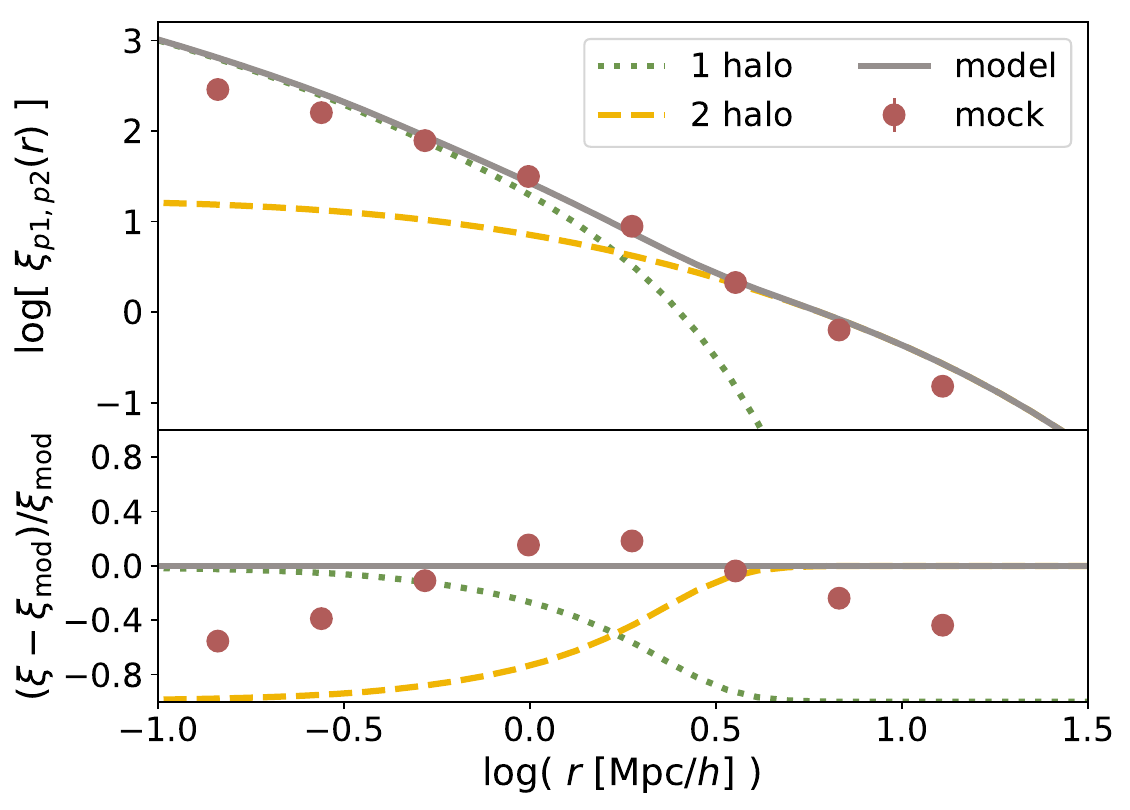}
    \caption{Comparison between the cross-correlation function, measured with the modified Landy-Szalay estimator of Eq.~\eqref{eq:cross_estimator}, between two dummy mock-populations at redshift $z = 0$. The lower panel shows the distance ratio between the measurement and the model prediction.}
    \label{fig:cross_corr}
\end{figure}

In Fig. \ref{fig:cross_corr} the red circles show the cross-correlation measured with Eq.~\eqref{eq:cross_estimator} for two different mock-populations with a dummy choice of the occupation probabilities parameters. 
Errors are measured using a bootstrap scheme with 10 sub-samples.
For comparison, we also show the halo-model prediction of the two-point correlation function, obtained with Eqs.~\ref{eq:hm_cross_1h} and \ref{eq:hm_cross_2h}, separated in 1-halo and 2-halo term contribution.
The lower panel of Fig.~\ref{fig:cross_corr} shows the distance ratio between the measure and the model prediction, which is lower than $40\%$ over almost all the scales inspected.


\section{Proof-of-concept application: ionizing photons production from Lyman-break galaxies at high redshift}
\label{sec:ioncase}

In the context of high redshift cosmology, one of the most compelling open problems is the process of Reionization, that brought the Universe from the optically thick state of the Dark Ages to the transparent state we observe today
\citep[see, e.g.,~][ for reviews]{ChoudhuryFerrara2006,Wise2019}.
Modelling this phase of the Universe evolution is a tricky task, especially using methods tuned to reproduce the observations at low redshift, such as hydrodynamical simulations and semi-analytical models. 
Instead, if we trust the capability of N-body simulations to capture the evolution of DM haloes up to the highest redshifts, an empirical method such as ScamPy is more likely to correctly predict the distribution of sources.

We expect Reionization to occur as a non-homogeneous process in which patches of the Universe ionize and then merge, prompted by the formation of the first luminous sources \citep{BarkanaLoeb2001}.
In order to map the spatial distribution of these ionized bubbles to the underlying dark matter distribution we apply our method to reproduce the observations of eligible candidates for the production of the required ionizing photon budget.
It is commonly accepted that the primary role in the production of ionizing photons at high redshift has been played by primordial, star forming galaxies \citep{shapiro1987,miralda-escude1990,BarkanaLoeb2001,steidel2001}.
The best candidates for these objects are Lyman-break galaxies (LBGs) which are selected in surveys using their differing appearance in several imaging filters, due to the position of the Lyman limit \citep{steidel2001,lapi2017,steidel2018,matthee2018}.

The first galaxies that started to inject ionizing radiation in the intergalactic medium were hosted in small DM haloes with masses up to a minimum of $10^8\ M_\odot/h$.
In order to paint a population of LBGs on top of a DM simulation, we need, first of all, a high resolution N-body simulation to provide the halo/sub-halo hierarchy required by ScamPy.
We therefore run the FoF and SUBFIND algorithms on top of 25 snapshots in the redshift range $4 \le z \le 10$ with thinness $\Delta z = 0.25$.
The DM snapshots have been obtained by running the (non-public) \textsc{P-GADGET-3} N-body code \citep[which is derived from the \textsc{GADGET-2} code,][]{gadget} on the two simulation dubbed \texttt{highres} and \texttt{midres} in Table~\ref{tab:cosmo_sim}.

Simulating the reionization process in more detail would required larger simulations at the same level of mass resolution as obtained for the \texttt{highres} and \texttt{midres} catalogues of our sample, the computational cost becoming out of reach for the test we want to perform here.
Overcoming the computational cost of N-body simulations could be obtained by applying up-sampling techniques of sub-grid modelling.
Such methods use a low resolution density field and build mock halo catalogues either by matching the theoretical predictions of the halo mass function \cite{delaTorrePeacock2013,Angulo2014} or the bias evolution in time \citep{nasirudin2019}.
This goes beyond the aim of the test-bench application we want to present here and we reserve to exploit it in future extensions of this work.

To set the occupation probabilities parameters with the likelihood in Eq.~\eqref{eq:loglike00}, we need the 1- and 2-point statistics of LBGs at high redshift.
The observational constrains of these high redshift statistics, due to the high distances involved, are not sufficiently tight.
We have therefore extrapolated the available measures as follows:
\begin{itemize}
    \item in \cite{Bouwens2019} the luminosity function of LBGs is fitted up to redshift $z=10$ and $M_\text{UV} \approx -16$. We extrapolate this fit up to $M_\text{UV} = -13$ and integrate to obtain an estimate of the number density of LBGs at high redshift. 
    \item \cite{Harikane2016} provide measurements of the angular 2-point correlation function in the redshift range $4 \le z \le 7$. We assume the clustering at redshift $z\ge7$ to be constant and equal to the measurement obtained for redshift $z=7$.
\end{itemize}
Both the observables assumed above are simplistic approximations which are though sufficient for test-benching our model.
We will investigate further the limits imposed by the lack of statistics at high redshift in future extensions of this work.

Once the parameters of the model have been set for each redshift, we run the algorithm described in Sec.~\ref{sec:liboverview} and obtain a set of LBG mock catalogues.
As a first approximation, let us define a neutral hydrogen distribution on top of each of our snapshots and consider the ionized region that should form around each source of our mock catalogues.
Each mock-LBG in our simulation is producing an amount of ionizing photons which is proportional to its UV luminosity, $M_\text{UV}$.
Namely, the rate of ionizing photons that escape from each UV source is 
\begin{equation}
    \label{eq_ion_photons}
    \dot{N}_\text{ion}(M_\text{UV}) \approx f_\text{esc}\; k_\text{ion}\; \text{SFR}(M_\text{UV})
\end{equation}
where $k_\text{ion} \approx 4 \times 10^{53}$ is the number of ionizing photons $\text{s}^{-1}(M_\odot/\text{yr})^{-1}$, with the quoted value appropriate for a Chabrier initial mass function (IMF), $f_\text{esc}$ is the average escape fraction for ionizing photons from the interstellar medium of high-redshift galaxies \citep[see, e.g.~][]{Mao2007,dunlop2013,furlanetto2015,lapi2017,chisholm2018,steidel2018,matthee2018}, and $\log( \text{SFR}(M_\text{UV}) ) \approx -7.4 - 0.4\;M_\text{UV}$ is the star formation rate of each source.
The volume of the Str\"omgren sphere, that forms around each mock-LBG, is then given by
\begin{equation}
    \label{eq:strom_vol}
    V_\text{S} \equiv \dfrac{\dot{N}_\text{ion}(M_\text{UV})}{\overline{n}_H(z)}\; t_\text{rec}\;(1 - e^{- t / t_\text{rec}})
\end{equation}
where $\overline{n}_H(z) \approx 2 \times 10^{-7}\;(\Omega_b(z) h^2/0.022)\ \text{cm}^{-3}$ is the mean comoving hydrogen number density at given redshift while $t$ is the cosmic time at given redshift and $t_\text{rec}$ is the cosmic time at the epoch the source started producing a steady flux of ionizing photons.

We make the following simplistic assumptions:
\begin{itemize}
    \item at each snapshot we do not provide any information about the previous reionization history: at each redshift sources have to completely ionize the medium and the value of $t_\text{rec}$ is fixed at the cosmic time corresponding to $z = 20$.
    \item the escape fraction is set to $f_\text{esc} = 0.1$, which is a conservative value with respect to what recent observations suggest.
\end{itemize}

\begin{figure*}
    \centering
    \includegraphics[width=0.9\textwidth]{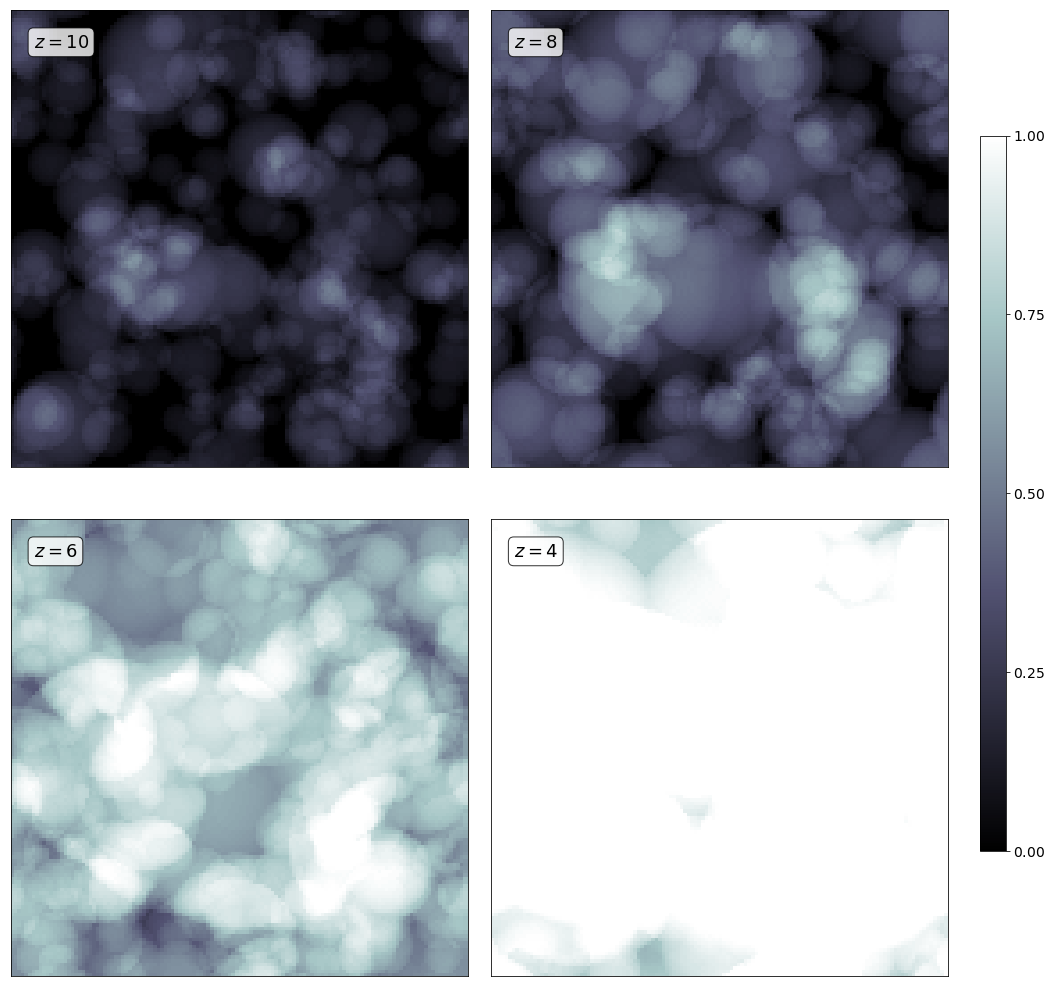}
    \caption{Four snapshots at different redshift (shown in the top left angle of each tile) of the ionization fraction obtained by projecting the values in each voxel along one dimension. The value of $X_\text{re}$ increases from darker to lighter shades of gray.}
    \label{fig:ion_slice_mosaic}
\end{figure*}

With the aforementioned simplifications, we can build spheres around each source at each redshift, therefore producing an approximated map of the ionization state of our snapshots, without having to rely on radiative transfer.
In Figure~\ref{fig:ion_slice_mosaic} we show the projection along one dimension of 4 snapshots at redshift $z = 10, 8, 6$ and $4$ for the \texttt{highres} simulation.
To get the point-by-point ionization fraction we divide our snapshots into voxels of fixed size. 
If a voxel is embedded within the Str\"omgren sphere belonging to some source, it is set as ionized, otherwise it is considered neutral.
Voxels that lie in the overlapping of two or more Str\"omgren spheres are counted only once.
This further approximation implies \textit{loosing} an amount of ionizing photons which is proportional to the overlapping volume of the whole simulation box.
This approximation mainly affects the lower redshifts, as shown in the next Section.
In Figure~\ref{fig:ion_slice_mosaic} we set the voxel-side to $0.25\ \text{Mpc}/h$, resulting in $128^3$ voxels in total.

In the remaining part of this Section we will show the results of some measurements that can be obtained from these mock ionization snapshots.

\subsection{Ionized fraction measurement}
\label{sec:ionfrac}

\begin{figure}
    \centering
    \includegraphics[width=0.45\textwidth]{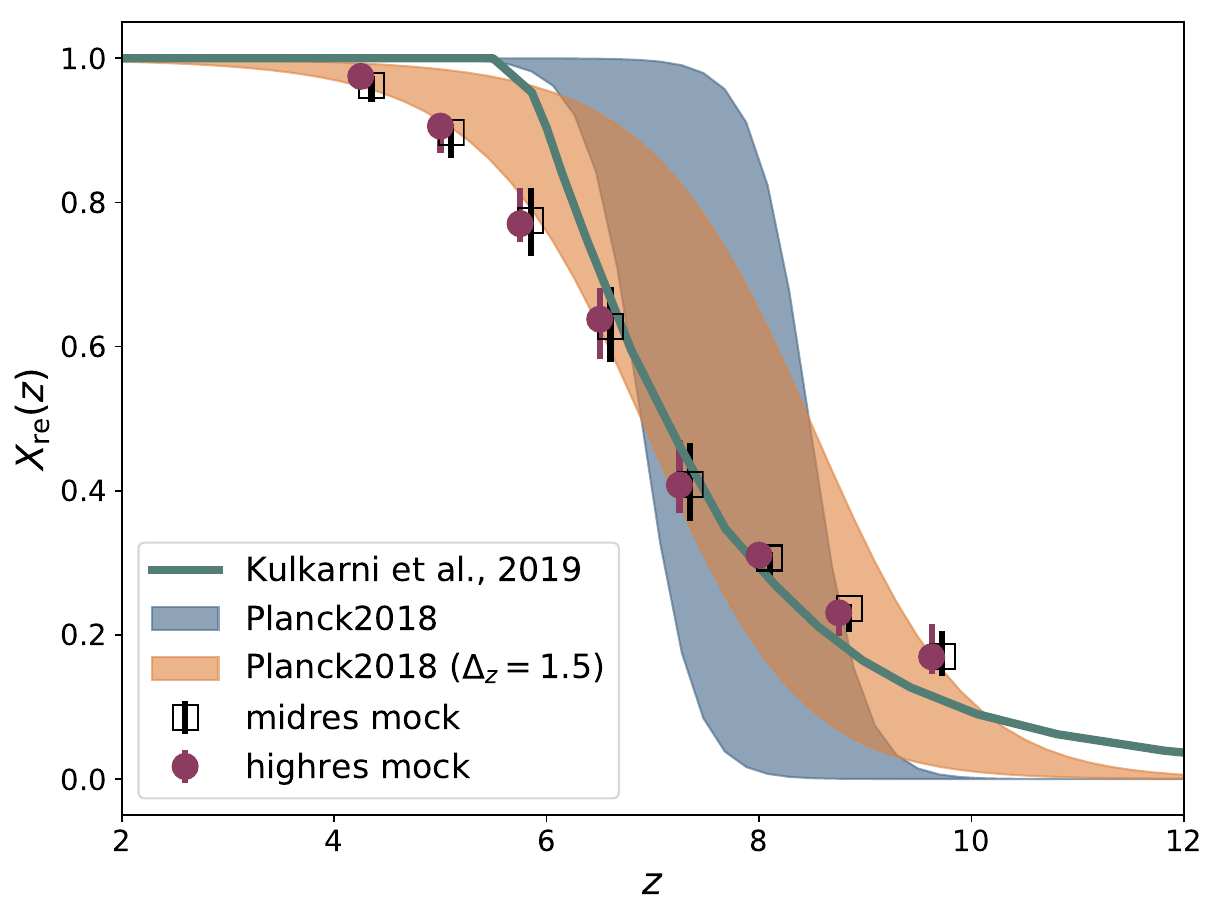}
    \caption{
    Evolution of the hydrogen ionization fraction $X_\text{re}$ with redshift. Red circles and empty squares mark the measurements obtained with our method on the \texttt{midres} and \texttt{highres} simulation, respectively. The \texttt{midres} distribution has been shifted with an offset of $\delta z = 0.1$ along the x-axis direction to better distinguish it from the \texttt{highres} one. The shaded regions delimit the model used in \protect\cite{Planck2018} and a modification for widening the reionization window. The solid line shows the prediction from \protect\cite{kulkarni2019}.
    }
    \label{fig:ionfrac}
\end{figure}

At each redshift, we measure the ionization fraction resulting from our pipeline by counting the number of voxels marked as ionized over the total number of voxels in which the snapshot is divided.
The results for both the \texttt{midres} and the \texttt{highres} simulations are marked with empty squares and red circles, respectively, in Figure~\ref{fig:ionfrac}.
We compare our measurement with models of reionization history from recent literature.

The gray shaded region shows the $\tanh$-model used in \cite{Planck2018} while the orange one delimits the prediction of the same model with a larger value of the parameter that regulates the steepness of the ionization fraction evolution \citep[$\Delta_z = 1.5$ instead of $\Delta_z = 0.5$, as from][]{Lewis2008}.
The solid green line shows the model from \cite{kulkarni2019} which is obtained by computing with the \textsc{ATON} code \citep{aubertteyssier2008,aubertteyssier2010} the radiative transfer a-posteriori on  top of a gas density distribution obtained using the \textsc{P-GADGET-3} code with the \textsc{QUICK\_LYALPHA} approximation from \cite{viel2004}.

Our mock ionization boxes predict reionization to reach half-completion ($X_\text{re} = 0.5$) at redshift $z = 6.88_{-0.13}^{+0.12}$, which is a lower value with respect to the \cite{Planck2018} prediction of $z = 7.68 \pm 0.79$, but still within the error bars.
Comparing to the extremely steep model used in \cite{Planck2018}, the evolution in our simulations is way shallower, closer to the lower limit of the modified $\tanh$-model.
Nonetheless, our measurements seem to agree fairly well with the measurements of \cite{kulkarni2019} up to redshift $z \approx 6$.
With respect to the other authors, our simulation reaches completeness (i.e.~$X_\text{re} = 1$) at redshift $z \approx 4$.
As anticipated, this issue at the lowest redshifts is by some extent expected.
The overall ionizing photon budget is indeed under-estimated in our approximation as the result of how we treat the overlapping region between different Str\"omgren spheres.
We plan to address this point, by implementing a physically motivated treatment of these overlapping regions \citep[as, e.g., in ][]{zahn2007} in future extensions of our analysis.

Taking into account the strong approximations made in this proof-of-concept application, the measurement we obtain for the evolution of $X_\text{re}$ is surprisingly consistent with equivalent measures in literature.

\subsection{Ionized bubble size distribution}
\label{sec:iondist}

It is accepted that reionization results from the percolation of ionized HII bubbles as well as from their growth in radius \citep{miralda-escude1990,furlanetto2004,WangHu2006} in the neutral intergalactic medium.
A relevant statistics for cosmological studies is the size distribution of the individual bubbles forming around ionizing radiation sources.
Obtaining precise measurements of this statistics could help constraining future experiments, such as CMB-S4 \citep{roy2018} and 21cm intensity mapping \citep[e.g.~][]{mesinger11}.

In our framework, getting estimates of the bubble size distribution is straightforward.
\begin{figure*}
    \centering
    \includegraphics[width=0.95\textwidth]{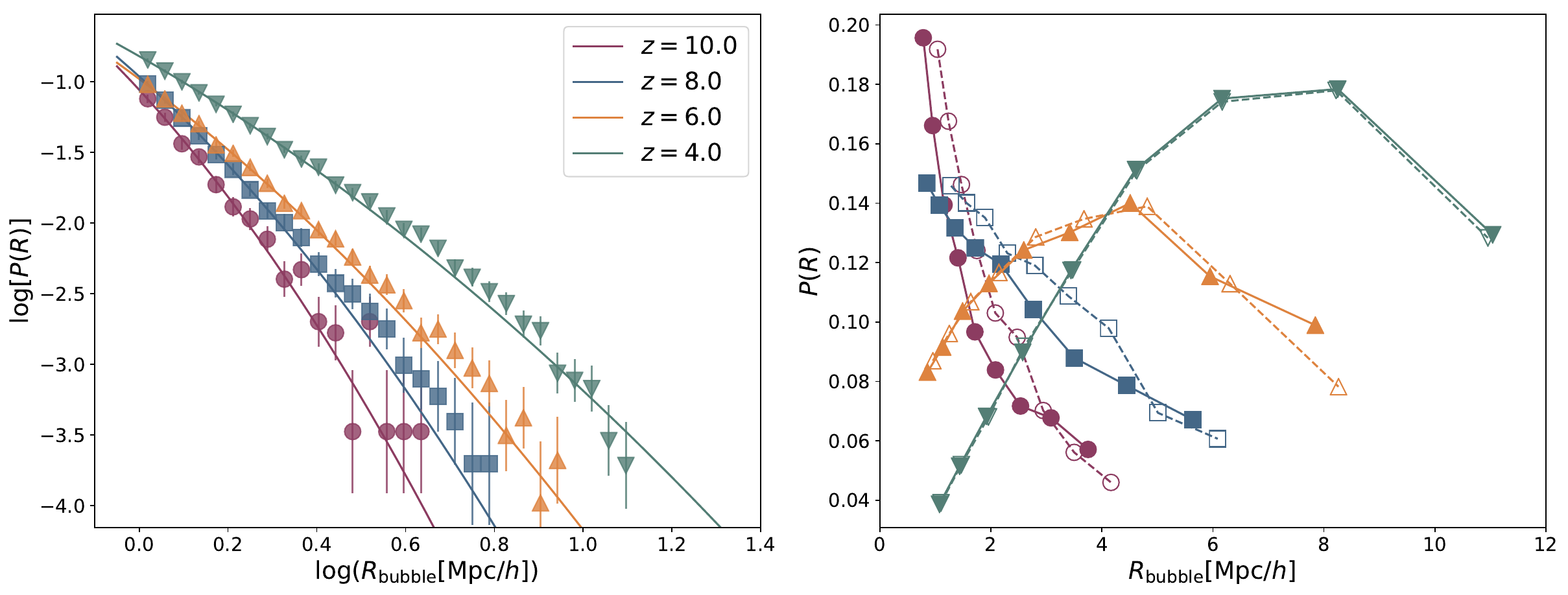}
    \caption{Bubble size probability distributions. \textit{Left panel:} fraction of the bubbles with given size over the total number of bubbles, markers are measured from the \texttt{highres} simulation while the solid lines show the model of Eq.~\protect\eqref{eq:bubble_sizeprob} fitted on these data (the best fitting parameters are listed in Table~\protect\ref{tab:bubble_size_par}). \textit{Right panel:} fraction of ionized voxels embedded in bubbles with given size over the total number of ionized voxels. Dashed lines mark the measurements obtained from the \texttt{midres} simulation, while solid lines have been obtained from the \texttt{highres} simulation.}
    \label{fig:bubble_prob}
\end{figure*}
In Figure~\ref{fig:bubble_prob} we present measurements of two different definitions for the bubble size probability.

On the left panel we plot the fraction of bubbles of given size over the total number of bubbles in the simulation box.
The measurement has been obtained at redshift $4 \le z \le 10$, with bin size $\delta z = 1$, we plot results only for $z = 4, 6, 8$ and $10$ for clarity.
The distribution shown presents a log-normal shape that we fit with the model from \cite{roy2018}
\begin{equation}
    \label{eq:bubble_sizeprob}
    P(R) = \dfrac{1}{R}\;\dfrac{1}{\sqrt{2\pi\sigma_{\ln r}^2}}\exp\biggl\{- \dfrac{\bigl[\ln(R/\overline{R})\bigr]^2}{2 \sigma_{\ln r}^2}\biggr\}\text{;}
\end{equation}
the model is regulated by two free parameters: the characteristic bubble size $\overline{R}$ (in $\text{Mpc}/h$) and the standard deviation $\sigma_{\ln r}$.
\begin{table}
    \centering
    \caption{Best-fit parameters of the log-normal model defined in Eq.~\protect\eqref{eq:bubble_sizeprob} obtained from our measures on the \texttt{highres} mock ionized bubble catalogue.}
    \label{tab:bubble_size_par}
    \begin{tabular}{ccc}
\toprule
$z$ & $\overline{R}$ $[\text{Mpc}/h]$ & $\sigma_{\ln r}^2$ \\ \midrule
10 & 0.229 $\pm$ 0.006 & 0.614 $\pm$ 0.025 \\
9 & 0.181 $\pm$ 0.006 & 0.786 $\pm$ 0.033 \\
8 & 0.222 $\pm$ 0.004 & 0.810 $\pm$ 0.024 \\
7 & 0.213 $\pm$ 0.003 & 0.974 $\pm$ 0.022 \\
6 & 0.165 $\pm$ 0.003 & 1.340 $\pm$ 0.033 \\
5 & 0.178 $\pm$ 0.003 & 1.616 $\pm$ 0.052 \\
4 & 0.208 $\pm$ 0.003 & 1.983 $\pm$ 0.052 \\
 \noalign{\smallskip}\hline \bottomrule
\end{tabular}
\end{table}
We list the best-fitting values of these parameters in Table~\ref{tab:bubble_size_par}, for the different redshifts considered.
While the value of the characteristic radius is almost constant in time with a value of $\overline{R} \approx 0.2\ \text{Mpc}/h$, the standard deviation increases significantly from higher to lower redshift.

On the right panel of Figure~\ref{fig:bubble_prob}, we show the fraction of ionized voxels as a function of the bubble radius over the total number of ionized voxels in the simulation box (normalized to $1$).
The solid lines show the measurements obtained from the \texttt{highres} box, while the dashed ones mark the distribution obtained from the \texttt{midres} box.
The results on the two boxes are consistent between the two simulations, especially at lower redshifts.
Compared to the left panel, the measurements obtained for the bubble size probability definition of the right panel are more consistent with what can be found in literature \citep[e.g.~][]{zahn2007}.
In particular, the characteristic radius seems to grow from higher to lower redshift, reaching values in the order of $1\div10\ \text{Mpc}/h$.
We could not fit the distributions on the right panel of Figure~\ref{fig:bubble_prob} with the same log-normal model of Eq.~\eqref{eq:bubble_sizeprob}.
This is probabily due to not having considered bubble overlapping in our measurements.
We will investigate further on this topic in future work.


\section{Summary \& Conclusions}
\label{sec:conclusions}

We have here presented ScamPy, our application for painting observed populations of objects on top of DM-only N-body cosmological simulations.
With the provided python interface, users can load and populate DM haloes and sub-haloes obtained by means of the FoF and SUBFIND algorithms applied to DM snapshots at any redshift.
We foresee to extend this framework to the usage with DM halo and sub-halo catalogues obtained with alternative algorithms.

The main requirements that guided the design of ScamPy were to provide a flexible and optimized framework for approaching a wide variety of problems, while keeping the computation fast and efficient.
To this end, we stick to the simple, yet physically robust, SCAM prescription for providing the recipe to populate DM haloes and sub-haloes.
In Section~\ref{sec:scam}, we presented an overview of the theoretical background of this methodology, while, in Section~\ref{sec:library} we provided a detailed description of the components and main algorithms implemented in ScamPy.

In Section~\ref{sec:v&v} we have shown a set of measurements obtained from simulations populated with galaxies using ScamPy.
We have demonstrated that the output mock-galaxies have the expected abundance and clustering properties. 
Furthermore, we have also proven that the same API could be used to ``paint'' multiple populations on top of the same DM simulation and that the cross-correlation between these populations is also well mimicked.

In the context of reionization, this could help in studying the spatial distribution properties and evolution in time of the ionized bubbles that might have developed around sources of ionizing radiation.
In Section~\ref{sec:ioncase}, we have performed a preliminary study, under simplistic assumptions, on the ionization properties of the high redshift Universe which result from the injection in the medium of ionizing photons from Lyman Break Galaxies (LBG).
We are now able to measure locally on simulations the ionized hydrogen filling factor at different redshifts.
This also allows to perform a tomographic measure of the ionization state of the medium at varying cosmic time.
Furthermore, we can also directly measure the ionized bubble size distribution, which is a quantity that, up to now, has been either modeled indirectly \citep{roy2018} or measured assuming radiative transfer \citep{zahn2007}.

While a specific problem prompted the development of the API, extensibility has been a crucial design choice. 
ScamPy features a modular structure exploiting Object-Oriented programming, both in \textsc{c++} and in python.
With a wise usage of polymorphism, we obtained a flexible application that can be both used by itself as well as along with other libraries.

We are working on adding to the API further miscellaneous functionalities, such as the possibility to download and install it with both \texttt{pip} and \texttt{conda}, in the next months. 
The online documentation is continuously updated, and more examples and tutorials are ready to be uploaded in the next months.

To conclude, we are planning to extend our work both on the scientific side, by exploring new directions, and on the computational side, by implementing an efficient $k^d$-tree algorithm for the optimisation of the neighbor search in both DM-halo catalogues and mock-galaxy catalogues.
A possible list of the directions we are planning to pursue follows.
\begin{itemize}
    \item {\emph{Application of the algorithm to multiple populations/different tracers of the LSS}}~-~ An application of the same pipeline to the other major players that are known to be involved on the reionization of hydrogen, e.g.~AGNs, would be trivial, provided that suitable datasets are available. Another possibility would be the cross-correlation with intensity mapping like e.g. \cite{spinelli19}.
    \item {\emph{Application to the reionization process}}~-~ Up to now we have mainly applied ScamPy functionalities to a proof-of-concept framework.
    Nonetheless, with the dataset at our disposal, an extensive study of the role played by LBG in reionization is possible, provided that larger N-body simulations with high resolution are available.
    This would require DM halo catalogues on large simulation boxes, with a complete sampling of the lower masses (down to $10^8 M_\odot$).
    Such catalogues could be obtained by running high resolution N-body simulations or, more realistically, by applying a sub-resolution scheme, such as the halo-bias models proposed by \cite{nasirudin2019}.
    \item {\emph{Extension to different cosmological models}}~-~ By now all our investigations have been performed assuming a $\Lambda$CDM cosmology. 
    To extend these results to other cosmological models would only require modest modifications of the source code and to obtain the corresponding DM-only N-body simulations, similar to high redshift studies performed
    e.g. in warm dark matter scenarios or massive neutrino cosmologies \cite{maioviel,fontaviel}.
    \item {\emph{Machine learning extension of the halo occupation model}}~-~ Using the halo occupation distribution (HOD) is straightforward and a lot of literature is available on the topic. This approach, though, comes with the limit that all the properties of the observed population have to be inferred from the mass of the host halo. This is known to be a rough approximation. 
    To overcome this limit, we plan to use, instead, a neural network (NN) model of the host halo occupation properties where the inputs of the NN are a set of known features of the halo/sub-halo hierarchy, such as the local environment around the halo and the dispersion velocity within the halo.
    Matter density based approaches for extending DM only N-body simulations have already been attempted.
    Recent efforts in this sense include work from \cite{siyu2019} which use NNs to predict the non-linear evolution of matter perturbations beyond the Zel'dovic Approximation.
    In another work,  
    \cite{yip2019}, paint baryons on top of simulations only run with DM.
    We plan instead to investigate the possibility to post process the baryonic effects on top of DM-only simulations in a halo-based framework.
\end{itemize}

\section*{Acknowledgements}

The Authors warmly thank the referee, Boryana Hadzhiyska, for the constructive report.
T.R. is thankful to Federico Marulli, Alfonso Veropalumbo and Luigi Danese for useful discussion.
This work has been partially supported by PRIN MIUR 2017 prot. 20173ML3WW 002, `Opening the ALMA window on the cosmic evolution of
gas, stars and supermassive black holes'. A.L. acknowledges the
MIUR grant `Finanziamento annuale individuale attivit\'a base
di ricerca' and the EU H2020-MSCA-ITN-2019 Project 860744
`BiD4BEST: Big Data applications for Black hole Evolution STudies'.
MV and TR are supported by the INFN-PD51 INDARK grant. MV is also
supported by financial contribution from the agreement ASI-INAF n.2017-14-H.0.

\section*{Data availability}

No new relevant data were generated or analysed in support of this research.
All the information for reproducing our results are available in the article and in the website of the presented package (\href{https://scampy.readthedocs.io}{scampy.readthedocs.io}).



\bibliographystyle{mnras}
\bibliography{main}



\appendix

\section{API structure \& bushido}
\label{sec:apdxbushido}

In the context of software development for scientific usage and, in general, whenever the development is intended for the use in Academia, the crucial aspects that would make the usage flexible are often overlooked.

In the development of ScamPy, we have considered the good practices in software development, such as cross-platform testing and the production of reasonable documentation for the components of the API.
We have outlined a strategy for keeping the software ordered and easy to read while maintaining efficient the computation.
The usage of advanced programming techniques, along with the design of a handy class dedicated to interpolation, also allowed to boost the performances of our code.

In this Appendix, we describe the framework we have developed, highlighting the best programming practices used, and commenting on the design choices.

The overall structure can be divided broadly into 4 main components:
\begin{itemize}
    \item \textbf{\textsc{c++} core} - it mainly deals with the most computationally expensive sections of the algorithm.
    \item \textbf{Python interface} - it provides the user interface and implements sections of the algorithm that do not need to be severely optimised.
    \item \textbf{Tests}, divided into \textbf{unit tests} and \textbf{integration tests}, are used for validation and consistency during code development.
    \item \textbf{Documentation}, provides the user with accessible information on the library's functionalities.
\end{itemize}
The organization of the source code is modular. 
Test and documentation sections are treated internally as modules of the library, and their development is, to some extent, independent to the rest of the API.
Furthermore, not being essential for the API operation, their build is optional.

The \href{https://mesonbuild.com/index.html}{Meson Build System} deals with compilation and installation of the library.
Much like the well-known CMake (\href{https://cmake.org/}{reference website}), it allows to ease the compilation and favours portability while automatizing the research and eventual download of external dependencies.

\subsection{Modularization}
\label{sec:apdxmodularization}

The \textsc{c++} and Python implementations are treated separately and have different modularization strategies.
As we already anticipated, the c++ language is adopted to exploit the performances of a compiled language.
Nonetheless, it also allows for multi-threading parallelisation on shared memory architectures.
This would not be normally possible in standard python because of the Global Interpreter Lock, which limits the processor to execute exactly only one thread at a time.

Each logical piece of the algorithm (see Section~\ref{sec:liboverview} and Figure~\ref{fig:alg_flowchart}) has been implemented in a different module.
This division has been maintained both in the core c++ implementation and in the python interface.
Bridging over the two languages has been obtained through the implementation of source \textsc{c++} code with a C-style interface enclosed in an \texttt{extern "C"} scope to produce shared-libraries with C-style mangling.
To wrap the compiled c++ libraries in python we use the CTypes module.
This choice was made because CTypes is part of the Python standard. 
Therefore no external libraries or packages are needed.
This choice favours portability and eases compilation.

All of the \textsc{c++} modules are organized in different sub-directories with similar structure:
\begin{itemize}
    \item \texttt{src} sub-directory, containing all the source files (\texttt{.cpp} extension);
    \item \texttt{include} sub-directory, containing all the header files (\texttt{.h} extension);
    \item a \texttt{meson.build} script for building.
\end{itemize}

All the Python implementation is hosted in a dedicated sub-directory of the repository.
Each module of the python interface to the API is coded in a separate file.
The python dependencies to the c++ implementation are included in the source files at compile time by the build system.

\begin{table}
    \centering
    \caption{Python modules of the API. The first column lists the module names and the second provides a short description of the module purpose. We divided the table in two blocks, separating the modules of the package that depend on the \textsc{c++} implementation from those that have a pure Python implementation.}
    \label{tab:py_modules}
    \begin{tabular}{ l p{5cm} }
    \toprule
    \textbf{Module} & \textbf{Purpose} \\ \hline \midrule
    \multicolumn{2}{c}{Wrapped from C-interface}\\ \midrule
    \texttt{interpolator} & Templated classes and functions for cubic-spline interpolation in linear and logarithmic space \\ \hline
    \texttt{cosmology} & Provides the interface and an implementation for cosmological computations that span from cosmographic to Power-Spectrum dependent functions, computations are boosted with interpolation \\ \hline
    \texttt{halo\_model} &  Provides classes for computing the halo-model derivation of non-linear cosmological statistics.\\ \hline
    \texttt{occupation\_p} & Provides the occupation probability functions implementation. \\ \hline \midrule
    \multicolumn{2}{c}{Python-only} \\ \midrule
    \texttt{objects} & Defines the objects that can be stored in the class \texttt{catalogue} of the \texttt{scampy} package, namely \texttt{host\_halo},  \texttt{halo} and \texttt{galaxy}. \\ \hline
    \texttt{gadget\_file} & Contains a class for reading the halo/sub-halo hierarchy from the outputs of the SUBFIND algorithm of GaDGET. \\ \hline
    \texttt{catalogue} & It provides a class for organizing a collection of host-haloes into an hierarchy of central and satellite haloes. It also provides functionalities for authomatic reading of input files and to populate the Dark Matter haloes with objects of type \texttt{galaxy}.\\ \hline
    \texttt{abundance\_matching} & Contains routines used for running the SHAM algorithm.\\ \hline
    \bottomrule
\end{tabular}
\end{table}
In Table~\ref{tab:py_modules}, we list all the Python-modules provided to the user.
They are divided between the \textsc{c++} wrapped and the Python only ones. 
All of them are part of the \texttt{scampy} package that users can import by adding a 
\begin{verbatim}
    /path/to/install_directory/python
\end{verbatim} 
to their \texttt{PYTHONPATH}.

\subsection{External dependencies}
\label{sec:apdxext_dep}

Scientific codes often severely depend on external libraries.
Even though a golden rule when programming, especially with a HPC intent, is to \emph{not reinvent the wheel}, external dependencies have to be treated carefully.
If the purpose of the programmer is to provide their software with a wide range of functionalities, while adopting external software where possible, the implementation can quickly become a \emph{dependency hell}.

For this reason, we decided to keep the dependence on external libraries to a reasonable minimum.
The leitmotiv being, trying not to be stuck on bottlenecks requiring us to import external libraries while maintaining the implementation open to the usage along with the most common scientific software used in our field.

The \texttt{c++} section of the API depends on the following external libraries:
\begin{itemize}
    \item \textbf{GNU Scientific Library} \citep[version $2$ or greater]{gsl_manual}: this library is widely used in the community and compiled binary packages are almost always available in HPC platforms. 
    \item \textbf{FFTLog} \citep{Hamilton2000}: also this library is a must in the cosmology community. In our API, we provide a \texttt{c++} wrap of the functions written in Fortran90. 
    We have developed a patch for the original implementation that allows to compile the project with Meson (see \href{https://github.com/TommasoRonconi/fftlog_patch}{fftlog\_patch on GitHub} for details).
    \item \textbf{OpenMP}: one of the most common APIs for multi-threading in shared memory architectures. It is already implemented in all the most common compilers, thus it does not burden on the user to include this dependency.
\end{itemize}

We are aware that a vast collection of libraries for cosmological calculations is already available to the community \citep{marulli2016,astropy2013,astropy2018}.
The intent of our \texttt{cosmology} module is not to substitute any of these but to provide an optimized set of functions integrated in the API without adding a further dependence on external libraries.
By using polymorphism (both static and dynamic) we tried to keep our API as much flexible as possible.
We explicitly decided to not force the dependence to any specific Boltzmann-solver to obtain the linear power spectrum of matter perturbations (see Section~\ref{sec:halomodel}), the choice is left to the user.

Furthermore, the choice of Python to build the user interface, allowed to easily implement functions that do not require any other specific library to work.
An example is the \texttt{abundance\_matching} module, which is almost completely independent to the rest of the API: the only other internal module needed is the \texttt{scampy.object} but all its functionalities can be obtained by using python lambdas and numpy arrays.

The only other python libraries used in ScamPy are:
\begin{itemize}
    \item \textbf{CTypes} which is part of the Python standard and is used for connecting the C-style binaries to the Python interface.
    \item \textbf{Numpy} which, despite not being part of the standard, is possibly the most common python library on Earth and provides a large number of highly optimized functions and classes for array manipulation and numerical calculations.
\end{itemize}

\subsection{Extensibility}
\label{sec:apdxextension}

Simplifying the addition of new features has been one of our objectives from the first phases of development.
We wanted to be able to expand the functionalities of the API, both on the \textsc{c++} side, in order to boost the performances, and on the Python side, in order to use the API for a wide range of cosmological applications.

This is easily achieved with the modular structure we have built up.
Adding a new \texttt{c++} module reduces to including a new set of headers and source files in a dedicated sub-directory.
Further details on the structure said sub-directory should have and on the way its build is integrated in the API will be provided in the library website.

Adding new modules to the Python interface is even simpler, as it only requires to add a new dedicated file in the \texttt{python/scampy} sub-directory.
Eventually, it can be also appended to the \texttt{\_\_all\_\_} list in the \texttt{python/scampy/\_\_init\_\_.py} file of the package.
In this case, it is not necessary to operate on the build system as it will automatically install the new module along with the already existing ones.

\section{Performances \& benchmarking}
\label{sec:apdxperfbench}

We have measured the performances of our API's main components and benchmarked the scaling and efficiencies of the computation at varying precision and work-load.
We will show here a set of time measurements performed on the two main components of the library: the \texttt{cosmology}~class and the \texttt{halo\_model}~class.
These are the two classes that would most affect the performances in real-life applications of our API.

\subsection{Wrapping benchmark}
\label{sec:apdxwrapbench}

First of all, in Tab.~\ref{tab:bench_wraps}, we show the execution time of the same function called from different languages.
Since our hybrid implementation requires to bridge through \textsc{c} to wrap in python the optimisations obtained in \textsc{c++}, it is interesting to compare their respective execution time.
Along with the python execution time, we also provide the ratio with respect to the reference \textsc{c++} time, $t_\textsc{c++}$, for the same function.
All the times are expressed in nanoseconds.

\begin{table}
    \centering
    \caption{Execution time in nanoseconds of the same function in different languages. For the python case, we also show the ratio with respect to the \textsc{c++} execution time. The timings reported are the average of 10 runs on the 4 physical cores with hyper-threading disabled of a laptop with Intel$^\circledR$ Core\texttrademark i7-7700HQ 2.80GHz CPU.}
    \label{tab:bench_wraps}
    \begin{tabular}{ l | c | c c }
    \toprule
    \textbf{Function} & \textbf{C++} & \textbf{Python} & $\mathbf{t_{\textbf{py}}/t_\textbf{C++}}$\\ \hline \midrule
    \multicolumn{4}{c}{\texttt{cosmology} class}\\ \midrule
	c.tor         &	2.520e+05	&	8.892e+05	&	3.528	\\ \hline
	$d_C( z )$    &	2.083e+03	&	5.984e+03	&	2.873	\\ \hline
	$n( M, z )$   &	1.186e+08	&	1.197e+08	&	1.009	\\ \hline \midrule
    \multicolumn{4}{c}{\texttt{halo\_model} class}\\ \midrule
	c.tor         &	3.011e+09	&	2.961e+09	&	0.983	\\ \hline
	$n_g( z )$    &	1.917e+04	&	2.851e+04	&	1.488	\\ \hline
	$\xi( r, z )$ &	3.396e+06	&	1.784e+06	&	0.525	\\ \hline
    \bottomrule
\end{tabular}
\end{table}
We are showing 3 typical member calls that are representative of the functionalities provided by the two classes.
For both of them, we measured the constructor time (c.tor), the time for executing a function that returns a scalar ($d_C(z)$ and $n_g(z)$) and the execution time for a function returning an array ($n(M, z)$ and $\xi(r,z)$).
It can be noticed that, especially for the \texttt{cosmology}~class, by calling the same function in python, the execution time increases.
The worst case is the \texttt{cosmology}~class constructor time that looses a factor $\sim 3.5$ in python.
It has to be noticed though, that the execution time is lower than a millisecond and, since the constructor is the member function that is called the less, this is not severely affecting the overall performance of the python interface.

Nonetheless, because of the larger number of function calls required by moving from one language to another, loosing some performance is expected.
What we did not expect is the gain in performance we are getting when moving to python, as it is shown in the last column of the \texttt{halo\_model}~class box of Tab.~\ref{tab:bench_wraps}.
This behaviour might be due to the different way memory is allocated, accessed and copied in python with respect to \textsc{c++}/\textsc{c}.
Moreover, the timers used for measuring the execution in the different languages are different.
Even by comparing measures taken with the same precision, it is not guaranteed to have the same accuracy.

\subsection{Halo-model performances}
\label{sec:apdxhmbench}

We have then tested the execution time of the \texttt{halo\_model}~constructor and member functions at varying work-load.
In our implementation, the \texttt{halo\_model}~class requires to define a set of interpolating functions at construction time, these functions can be defined using our \texttt{interpolator}~class (see Table~\ref{tab:py_modules}).
The interpolation accuracy depends on the resolution of the interpolation grid.
In the \texttt{halo\_model}~class, at fixed limits of the interpolation interval, this is controlled by the \emph{thinness} input parameter, which takes typical values $50\div200$ in real-life applications.

\begin{figure}
    \centering
    \includegraphics[width=0.45\textwidth]{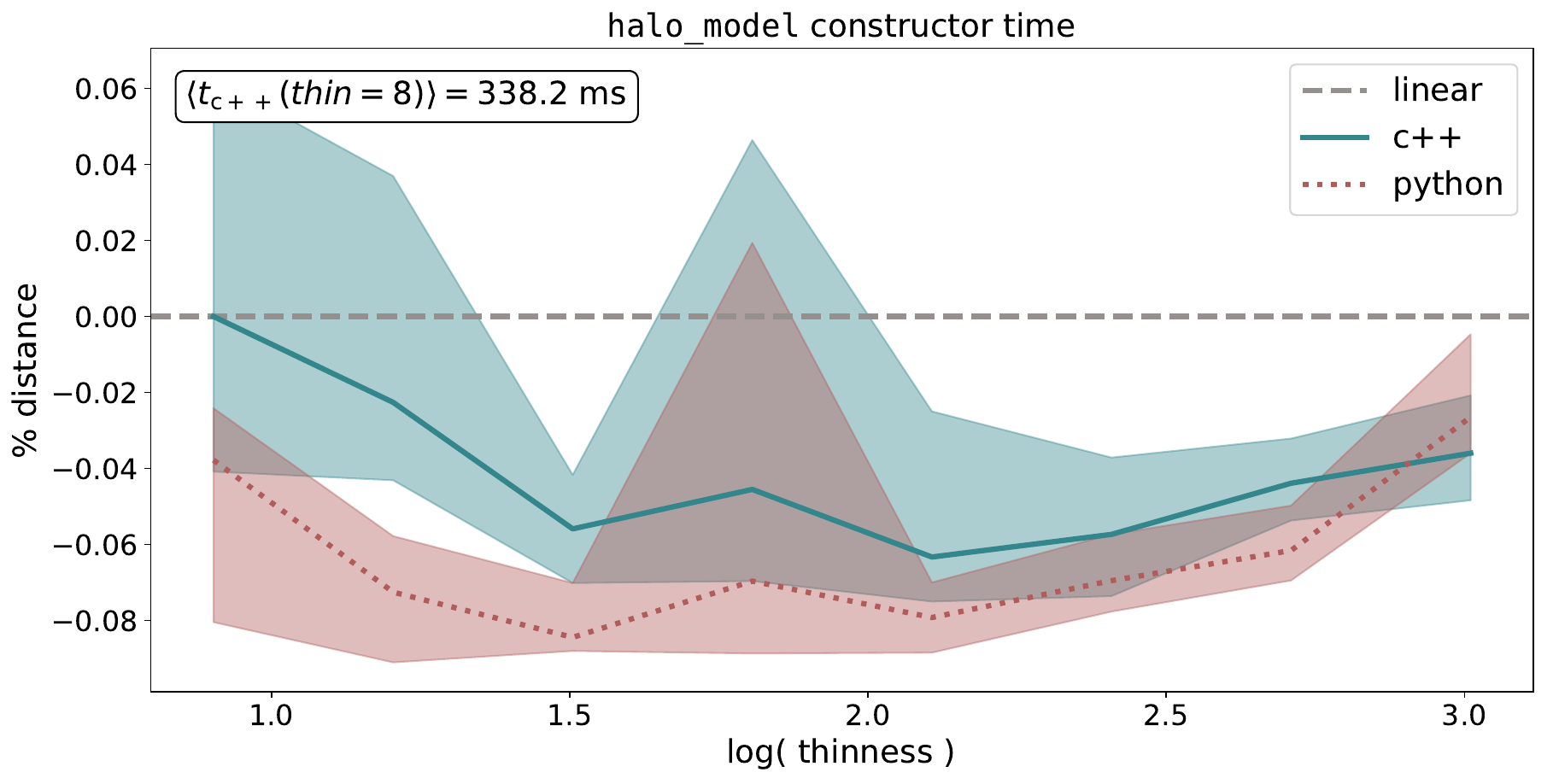}
    \caption{Percent distance between the constructor time scaling at varying thinness and the linear scaling case. For the python case, the percentage is computed with respect to the \textsc{c++} time to ease the comparison. For reference, we also show in the white text-box the measured constructor time with thinness $= 8$.}
    \label{fig:hm_constr_thinness}
\end{figure}

In Fig.~\ref{fig:hm_constr_thinness} we show how the constructor-time varies with varying thinness in the range $10 < thin < 10^3$.
The plot is obtained by calling 10 times the constructor per each thinness value and then averaging (solid and dotted lines).
The shaded region marks the best and worst execution time among the 10 runs.
Instead of the actual execution time we show the percent distance with respect to perfect linear scaling (dashed line) for both the \textsc{c++} case (blue) and the python case (red).
We define the percent distance at given thinness as
\begin{equation}
    \label{eq:perc_dist}
    \%\ \text{distance}(thin) \equiv 100 \cdot \dfrac{t(thin) - t_\text{lin}(thin)}{t_\text{lin}(thin)}
\end{equation}
where $t_\text{lin}(thin)$ is the execution time for given thinness in the linear scaling case, computed with respect to the \textsc{c++} case.
In the white text box of Fig.~\ref{fig:hm_constr_thinness} we also show the \textsc{c++} constructor time for $thin = 8$, as a reference.
As the picture shows, the scaling is almost perfectly linear, with a maximum distance of the $0.06\%$ in the \textsc{c++} case.

Possibly the most crucial computational bottleneck of the whole API is the time taken by the computation of a full-model.
With the term ``full-model'' we mean the execution of the two functions for computing the halo-model estimate of the 1- and 2-point statistics, namely $n_g(z)$ and $\xi(r, z)$.
In an MCMC framework, while the constructor is called only once, these two functions are called tens of thousands of times.
This is a necessary step to set the parameterisation of the SCAM algorithm.

As also shown in Tab.~\ref{tab:bench_wraps}, the execution of the two single functions takes an amount of time which is in the order of the millisecond in the \textsc{c++} case.
We can also notice that the execution time of a full-model is dominated by the computation of the two point correlation function, $\xi(r,z)$.
Since this function is operating on a vector and returning a vector, it is reasonable to expect that its execution time varies with the work-load, i.e.~with the vector size.

\begin{figure}
    \centering
    \includegraphics[width=0.45\textwidth]{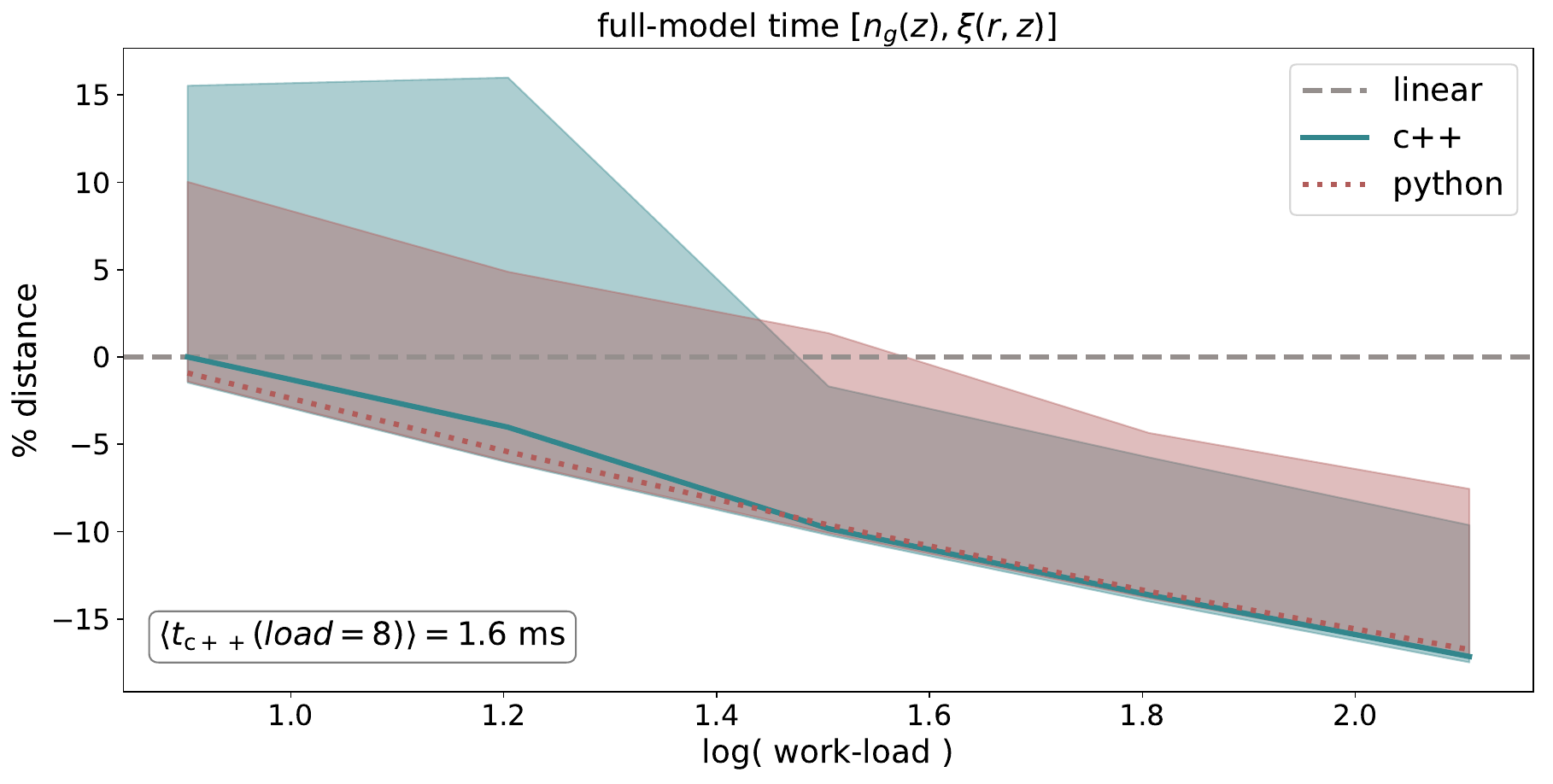}
    \caption{Percent distance between the full-model time scaling at varying work-load and the linear scaling case. For the python case, the percentage is computed with respect to the \textsc{c++} time to ease the comparison. For reference, we also show in the white text-box the measured constructor time with work-load $= 8$.}
    \label{fig:hm_fullmod_workload}
\end{figure}

In Fig.~\ref{fig:hm_fullmod_workload} we show the percent distance, defined as in Eq.~\eqref{eq:perc_dist}, of the average full-model execution time at varying work-load (solid lines) with respect to the perfect linear scaling case (dashed line), in the range $2^3 \le \text{load} \le 2^{14}$.
The measurements are obtained by averaging the results of 10 runs in both \textsc{c++} (blue) and python (red).
The shaded regions mark the best and worst performance among all the runs at varying workload.
It can be noticed that, by increasing the work-load, the average execution time gets up to $15\%$ worse than perfect linear scaling.
This is due to some latency introduced by the necessity of Fourier transforming the power spectrum to model clustering.
We have to point out though, that the typical work-load is in the range $5\div15$ for real-life applications and that the execution time in this cases is of the order of the millisecond.

Finally, we have measured how the constructor time scales with increasing number of multi-threading processors. 
We did not perform this measure for the full-model computation because, in the perspective of using it in a MCMC framework with parallel walkers, the full-model will be computed always serially.

We present measurements of both the constructor time \emph{strong scaling} and \emph{weak-scaling}.
While the first measures the scaling with processor number at fixed thinness, the latter measures the scaling at thinness increasing proportionally with the processor number.

First of all, let us define the \emph{speed-up}
\begin{equation}
    \label{eq:pb_sp00}
    S(p) = \dfrac{t( 1 )}{t( p )}
\end{equation}
where $p$ is the number of processors and $t(p)$ is the time elapsed running the code on $p$ processors.
This quantity measures the gain in performances one should expect when having access to larger parallel systems.

We also define the efficiency for the strong and the weak scaling case:
\begin{equation}
    \label{eq:pb_ep00}
    \begin{split}
        E_\text{strong}(p) &= \dfrac{S( p )}{p} \\
        E_\text{weak}(p) &= S( p ) \\
    \end{split}
\end{equation}
This quantity roughly measures the percentage of exploitation of the parallel system used.
Thus, providing a hint of how much the serial part of the code is affecting the gain we can expect from spawning multiple threads.

We run these measures on a node from the \texttt{regular}~partition of the SISSA Ulysses cluster.\footnote{Please refer to the \href{https://www.itcs.sissa.it/services/computing/hpc}{website} for detailed informations.}
Each of these nodes provide two shared memory sockets with 10 processors each.
We measured the constructor time by averaging the results of 100 runs where the threads number has been controlled by setting
\begin{verbatim}
    export OMP_NUM_THREADS=$ii
    export OMP_PLACES=cores
    export OMP_PROC_BIND=close
\end{verbatim}
where \texttt{ii} varies in the set $\{1, 2, 4, 8, 16, 20\}$ and where the last two commands control the affinity of the processes spawned.

\begin{figure}
    \centering
    \includegraphics[width=0.45\textwidth]{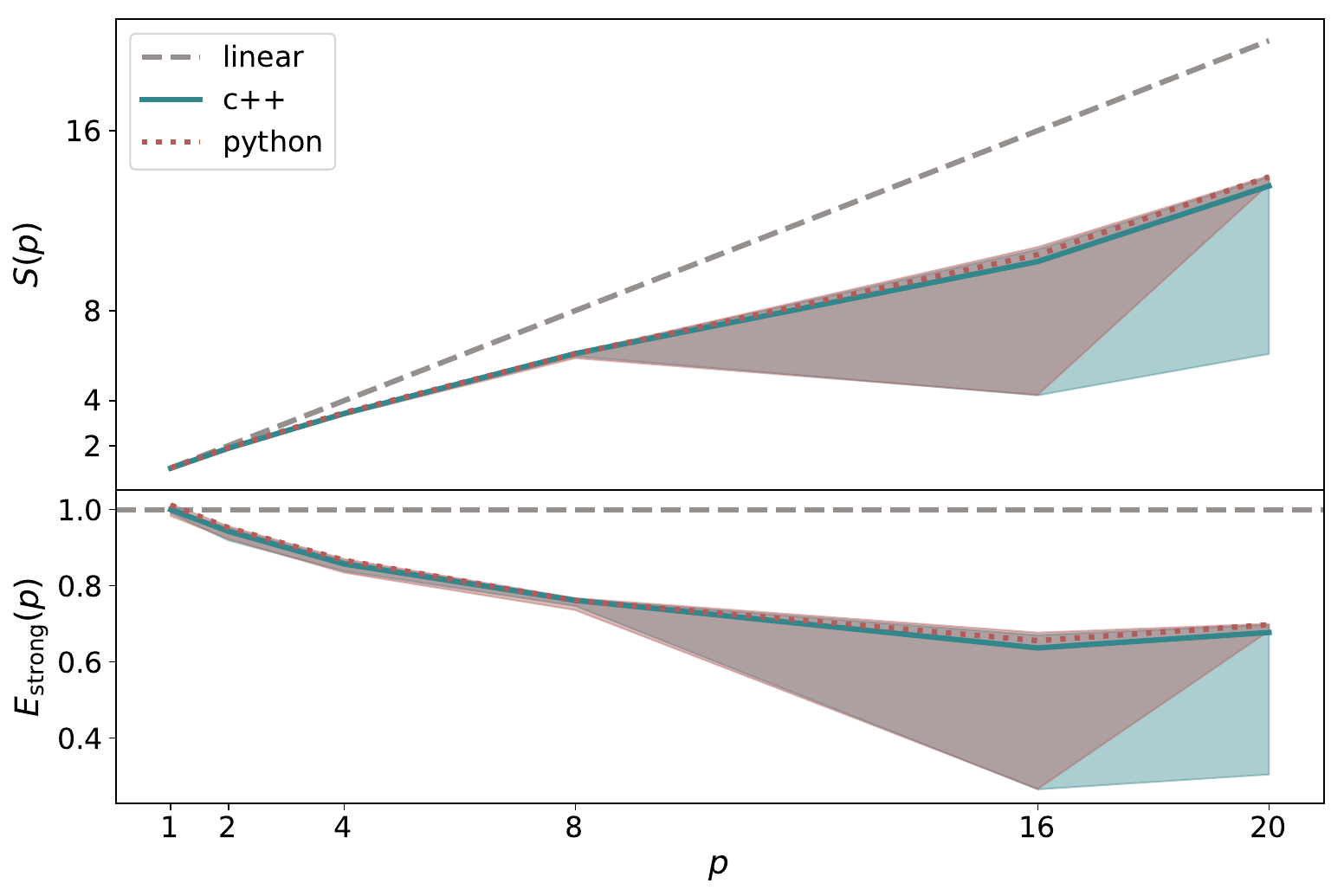}
    \caption{Strong-scaling speed-up (\emph{upper~panel}) and efficiency (\emph{lower-panel}) of the constructor time at fixed thinness and varying number of multi-threading processors. The solid line marks the average of 100 runs while the shaded region marks the best and worst result area. We run the tests on a full computing-node of the SISSA Ulysses cluster.}
    \label{fig:hm_constr_strong}
\end{figure}
In Fig.~\ref{fig:hm_constr_strong} we show the speed-up (upper panel) and efficiency (lower panel) of the strong scaling.
The dashed line marks perfect linear speed-up in the upper panel, and $100\%$ efficiency in the lower panel.
Even though it is far from being perfect, the speed-up shows a constantly increasing trend.
The efficiency seams to get constant around the $60\%$ for $p \ge 16$, but a larger parallel system would be necessary for getting a more precise measurement.

\begin{figure}
    \centering
    \includegraphics[width=0.45\textwidth]{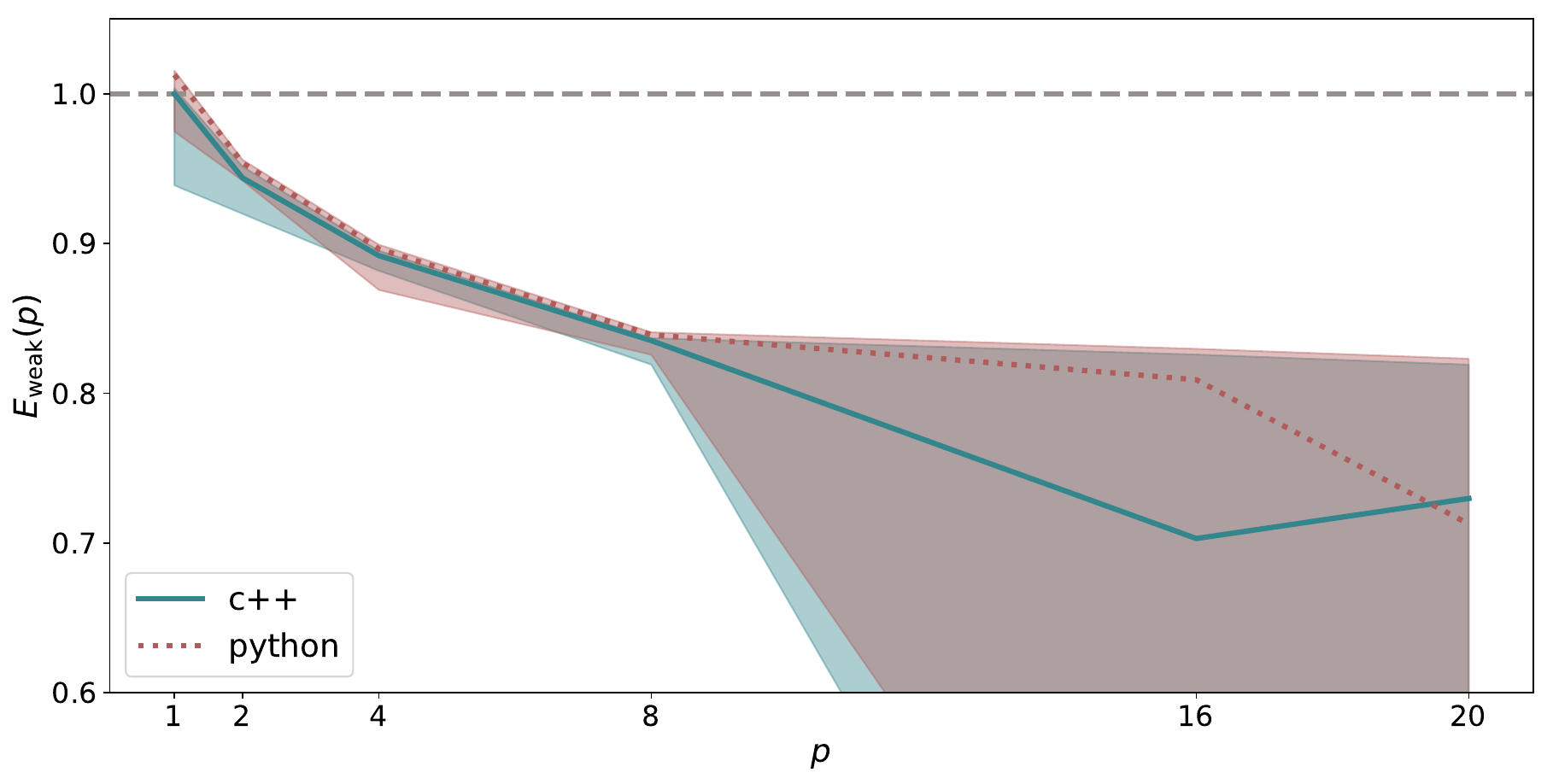}
    \caption{Weak-scaling efficiency of the constructor time at thinness growing proportionally to the number of multi-threading processors. The solid line marks the average of 100 runs while the shaded region marks the best and worst result area. We run the tests on a full computing-node of the SISSA Ulysses cluster.}
    \label{fig:hm_constr_weak}
\end{figure}
To conclude, in Fig.~\ref{fig:hm_constr_weak}, we show the the weak scaling efficiency case.
The thinness, at given processors number $p$, is set to $thin = 50 \cdot p$.
As the picture shows, the efficiency seams to become almost constant at $p \gtrsim 8$ for both the \textsc{c++} and python case, with a value between $70\%$ and $80\%$.


\bsp	
\label{lastpage}
\end{document}